\documentclass[%
 reprint,
superscriptaddress,
 amsmath,amssymb,
 aps,
prb,
]{revtex4-2}

\usepackage{graphicx}
\usepackage{dcolumn}
\usepackage{bm}

\usepackage[utf8]{inputenc}
\usepackage[T1]{fontenc}
\usepackage{mathptmx}
\usepackage{siunitx}
\usepackage{amsmath}
\usepackage{lipsum}
\usepackage{array}
\usepackage[normalem]{ulem} 
\usepackage{xcolor}
\usepackage[ colorlinks,
    linkcolor={blue},
    citecolor={blue},
    urlcolor={blue}]{hyperref}
\usepackage{braket}
\usepackage{float}

\newcommand{\mathd}{\mathrm{d}}

\newcommand{\tmop}[1]{\ensuremath{\operatorname{#1}}}
\newcommand\eqt{\hspace{0.17em}{=}\hspace{0.17em}}
\newcommand\equivt{\hspace{0.17em}{\equiv}\hspace{0.17em}}
\newcommand\intext{\hspace{0.17em}{\in}\hspace{0.17em}}

\newcommand\pt{\hspace{0.17em}{+}\hspace{0.17em}}
\newcommand\mt{\hspace{0.17em}{-}\hspace{0.17em}}
\newcommand\cdott{\hspace{0.12em}{\cdot}\hspace{0.12em}}

\newcommand{\br}{\mathbf{r}}
\newcommand{\bP}{\mathbf{P}}

\newcommand{\bQ}{\mathbf{Q}}
\newcommand{\bG}{\mathbf{G}}
\newcommand{\bk}{\mathbf{k}}
\newcommand{\bR}{\mathbf{R}}
\newcommand{\bS}{\mathbf{S}}
\newcommand{\bM}{\mathbf{M}}

\newcommand{\kl}{\bk_\ell}
\newcommand{\emiklR}{e^{-i\kl\cdot\bR}}
\newcommand\simt{\hspace{0.17em}{\sim}\hspace{0.17em}}
\newcommand{\rightarrowtext}{\hspace{0.1em}{\rightarrow}\hspace{0.1em}}
  \newcommand\timest{\hspace{0.09em}{\times}\hspace{0.06em}}
\newcommand{\bq}{\mathbf{q}}
\newcommand\kt{\hspace{0.17em}{<}\hspace{0.17em}}
\newcommand\gt{\hspace{0.17em}{>}\hspace{0.17em}}
\newcommand\rc {r_\text{c}}

\newcommand{\Vtworr}{V_{\rc}}
\newcommand{\Vtbvk}{V_{\rbvk}(\br,\br')}
\newcommand{\rbvk}{r_\text{SC}}

\newcommand{\Vt}{\Vtworr(\br-\br')}
\newcommand{\emikR}{e^{-i\mathbf{k}\cdot\mathbf{R}}}
\newcommand{\eikR}{e^{i\mathbf{k}\cdot\mathbf{R}}}
\newcommand{\bo}{\mathbf{0}}
\newcommand{\intbz}{ \int\limits_\text{BZ} \frac{d\mathbf{k}}{\Omega_\text{BZ}}}
\newcommand{\intbzq}{ \int\limits_\text{BZ} \frac{d\mathbf{q}}{\Omega_\text{BZ}}}

\newcommand{\hks}{h_{\mu\nu} (\bk)}
\newcommand{\bb}{\mathbf{b}}
\newcommand{\ba}{\mathbf{a}}
\newcommand{\bT}{\mathbf{T}}
 \newcommand{\sik}{  \sum_{i\bk}^{\text{occ}}}
 \newcommand{\sak}{  \sum_{a\bk}^{\text{empty}}}
\newcommand{\pik}{\psi_{i\bk} }

\newcommand{\pikrp}{\pik(\br')}
\newcommand{\pak}{\psi_{a\bk}}
\newcommand{\pakr}{\pak(\br)}

\newcommand{\expepsik}{ e^{-|(\varepsilon_{i\bk}^\text{DFT}-\varepsilon_\text{F}) \tau|}}
\newcommand{\expepsak}{e^{ -|(\varepsilon_{a\bk}^\text{DFT}-\varepsilon_\text{F}) \tau|}}
\newcommand{\gwsoc}{{{G_0W_0}\text{+SOC}}}
\newcommand{\res}{\mathfrak{R}}
\newcommand{\intcell}{\int\limits_\text{cell} d\br}
\newcommand{\bttt}{\bT_{t_1,\,t_2}}

\definecolor{darkgreen}{rgb}{0.0,0.55,0.0}
\newcommand{\RP}[1]{{\color{black} #1}} 
 
\newcommand{\DHP}[1]{{\color{black} #1}} 
\newcommand{\ASH}[1]{{\color{black} #1}}
\newcommand{\reviewnew}[1]{{\color{black} #1}}

\begin{document}

\preprint{APS/123-QED}



\title{ Efficient \textit{GW} band structure calculations using Gaussian basis functions and application to atomically thin transition-metal dichalcogenides}

\author{Rémi Pasquier}\email{remi.pasquier@physik.uni-regensburg.de}
\affiliation{Institute of Theoretical Physics and Regensburg Center for Ultrafast Nanoscopy (RUN), University of Regensburg, 93053 Regensburg, Germany}

\author{Mar\'ia Camarasa-G\'omez}
\affiliation{Centro de F\'isica de Materiales (CFM-MPC), CSIC-UPV/EHU,
 Paseo Manuel de Lardizabal 5, 20018 Donostia-San Sebasti\'an, Spain}
 \affiliation{Departamento de Pol\'imeros y Materiales Avanzados: F\'isica, Qu\'imica y Tecnolog\'ia,
Facultad de Qu\'imica, UPV/EHU, 20018 Donostia-San Sebasti\'an, Spain}

\author{Anna-Sophia Hehn}
\affiliation{Institute of Physical Chemistry, Christian-Albrechts-University Kiel, Max-Eyth-Strasse 1, 24118 Kiel, Germany
}

 \author{Daniel \surname{Hernang\'{o}mez-P\'{e}rez}}
 \affiliation{CIC nanoGUNE BRTA, Tolosa Hiribidea 76, 20018 San Sebasti\'an, Spain}

\author{Jan Wilhelm}\email{jan.wilhelm@physik.uni-regensburg.de}
\affiliation{Institute of Theoretical Physics and Regensburg Center for Ultrafast Nanoscopy (RUN), University of Regensburg, 93053 Regensburg, Germany}

\date{\today}

\begin{abstract}
We present a \textit{GW} space-time algorithm for periodic systems in a Gaussian basis  including spin-orbit coupling. 
We employ lattice summation to compute the irreducible density response and the self-energy, while we employ $k$-point sampling for computing the screened Coulomb interaction. 
Our algorithm enables accurate and computationally efficient quasiparticle band structure calculations for atomically thin transition-metal dichalcogenides. 
For monolayer MoS$_\text{2}$, MoSe$_\text{2}$, WS$_\text{2}$, and WSe$_\text{2}$, computed \textit{GW} band gaps agree on average within 50~meV with plane-wave-based reference calculations.
$G_0W_0$ band structures are obtained in less than two days on a laptop (Intel i5, 192 GB RAM) or in less than 30 minutes using 1024 cores. 
Overall, our work  provides an efficient and scalable framework for \textit{GW} calculations on atomically thin materials.
\end{abstract}

\maketitle

\section{Introduction}

\textit{GW} calculations~\cite{hedin1965new,Reining2018,Golze2019} have become a standard  method for calculating electron addition and removal energies of molecules~\cite{Blase2011,Faber2011,gw100,Knight2016}, two-dimensional materials~\cite{Molina2013,Gjerding2021,Altman2024} and bulk solids~\cite{bruneval2008accurate,Klimes2014,Leppert2024}.  
 Recent advancements of the $GW$ method span a broad spectrum,  including the application to deep core excitations~\cite{Golze2018,Golze2020,keller2020relativistic,mejia2021,Mejia2022basis,li2022benchmark,panades2023,Galleni2024,Kocklaeuner2025}, relativistic \textit{GW} schemes~\cite{holzer2019ionized,Yeh2022,foerster2023twocomponent,kehry2023,Gaurav2024,Abraham2024} and vertex corrections~\cite{Grueneis2014,Maggio2017Vertex,Lewis2019,Tal2021,wangy2021,Foerster2022ExploringG3W2,Lorin2023,bruneval2024fully,Wen2024,Foerster2025}.
These developments have firmly established \textit{GW} as a powerful and versatile approach within the domain of many-body perturbation theory.

Despite the methodological maturity of $GW$, computational challenges remain, particularly when applied to atomically thin materials. 
One of the primary limitations arises from the use of plane-wave basis sets in systems with vacuum, such as molecules or low-dimensional systems.
The need to represent the vacuum leads to large plane-wave basis set size and significant computational cost. 
These constraints motivate the development of alternative basis representations. 
One compelling solution involves atom-centered basis functions, which are localized and naturally adapted to such geometries. 
Atom-centered basis functions are already the standard for $GW$ implementations targeting molecules~\cite{Blase2011,gw100,Knight2016,Golze2019,Faber2011}. 
Also, several implementations of $GW$ with atom-centered basis functions and periodic boundary conditions have been reported~\cite{Strinati1980,Strinati1982,Rohlfing1993,Rohlfing1995,Wilhelm2017,zhu2021all,Ren/etal:2021,Harsha2024,Graml2024}. \ASH{
The first pioneering $GW$ calculations in semiconductors were carried out by Strinati~\textit{et al.}~\cite{Strinati1980,Strinati1982}, who studied the band structure of diamond using a local-orbital formulation of the $GW$ calculation.
In the 90ies,  Rohlfing~\textit{et al.}~\cite{Rohlfing1993,Rohlfing1995} studied bulk semiconductors and a silicon surface finding good agreement to plane-wave based $GW$ implementations in the order of 0.1~eV or better. 
The more recent periodic $GW$ implementations with atom-centered basis functions~\cite{Wilhelm2017,zhu2021all,Ren/etal:2021,Harsha2024} focus on 3D crystals and report similar precision. 
The $GW$ implementations~\cite{Strinati1980,Strinati1982,Rohlfing1993,Rohlfing1995,Wilhelm2017,zhu2021all,Ren/etal:2021,Harsha2024} use a formulation of $GW$ in frequency and rely on $k$-point sampling to account for the periodic boundary conditions. 
An alternative is the $GW$ space-time method~\cite{rojas_1995,liu2016} which has been originally formulated using plane waves and real-space representations as well as time and frequency representations.  
The $GW$ space-time method can be also reformulated using an atom-centered basis set~\cite{wilhelm2018,foerster2020,duchemin2021,foerster2022quasiparticle} which eliminates the use of plane-waves and real-space grids and allows for low-scaling $GW$ calculations on large molecules.
The $GW$ space-time approach in an atom-centered basis can be also combined with periodic boundary conditions, as we have demonstrated in our previous work~\cite{Graml2024} where we employed a $\Gamma$-only approach for the density response function and the self-energy, while relying on dense $k$-point sampling for the screened Coulomb interaction~$W$ to account for the divergence of $W$ at the $\Gamma$-point.
This implementation enabled the study of a twisted transition-metal dichalcogenide heterobilayers with almost 1000 atoms in the unit cell. 
The drawback of the $\Gamma$-only implementation~\cite{Graml2024} is that large unit cells are required where the density response needs to vanish on the length scale of the unit cell. 
In this work, we address the limitation of the $\Gamma$-point only approach~\cite{Graml2024}, which requires large unit cells, by introducing a lattice summation over neighbor cells for both the density response and self-energy, as  originally proposed in the $GW$ space-time method~\cite{rojas_1995}.
This extension enables the accurate and efficient treatment of crystals with small unit cells. 
The resulting $GW$ algorithm is particularly well-suited for low-dimensional materials, as the number of neighboring cells required in the lattice sums is significantly reduced compared to 3D bulk crystals. 
Furthermore, the   Gaussian basis set is efficient in simulations involving large vacuum regions, a common requirement when modeling low-dimensional systems.
}
Our implementation also supports the inclusion of relativistic effects via spin-orbit coupling (SOC) from Gaussian dual-space pseudopotentials~\cite{Hartwigsen1998,Krack2005,Vogt2025} and a perturbative correction to the quasiparticle energies.
We focus on $GW$ band structure calculations of atomically thin transition metal dichalcogenides (TMDCs) MoS$_2$, MoSe$_2$, WS$_2$, WSe$_2$ and we benchmark our $GW$ band structures against \DHP{state-of-the-art} plane-wave-based $GW$ implementations in BerkeleyGW~\cite{Deslippe2012} and VASP~\cite{Klimes2014}. 
We demonstrate that our $GW$ algorithm yields accurate and converged quasiparticle band structures across multiple convergence parameters, including basis set size, $k$-point sampling, the summation of neighbor cells, and the time- and frequency-mesh. 
We also discuss the computational efficiency of our $GW$ band structure algorithm, which enables a $GW$ band structure calculation of an atomically thin material on a laptop (Intel i5 13600HX, 192 GB RAM) within roughly a day.

\section{\textit{GW} space-time algorithm}\label{sec2}

Many efficient $GW$ algorithms~\cite{liu2016,foerster2020,duchemin2021} build on the $GW$ space-time method~\cite{rojas_1995}.
In order to introduce the basic idea of the $GW$ space-time method, we use a generic formulation in this section for non-periodic systems projecting all quantities on real-space grids. 
It is important to
note that this formulation differs from the original $GW$ space-time method~\cite{rojas_1995} where some
quantities were calculated using a plane-wave basis set.

In this work, we employ the $G_0W_0$ scheme which starts from a self-consistent {Kohn-Sham density functional theory (KS-DFT)} calculation~\cite{Kohn1965},
\begin{equation}
[h_0 (\br) + v^\text{xc} (\br)] \psi_n (\br) =
   \varepsilon_n^{\tmop{DFT}} \psi_n (\br)\,. \label{e1}
\end{equation}
$h_0$ contains the kinetic energy, the Hartree potential and
the external potential, while the exchange-correlation potential~$v^\text{xc}$ contains all electron-electron interactions beyond Hartree. $\psi_n(\br)$ is the KS orbital~$n$ and $\varepsilon_n^{\tmop{DFT}}$ the associated KS eigenvalue.
The terms $G_0$ and $W_0$ indicate that the Green's function~$G$ and the screened Coulomb interaction~$W$ are both computed using KS orbitals and KS eigenvalues, i.e., self-consistent updates of $G$ and $W$ from Green's function theory are omitted in $G_0W_0$. 

KS orbitals and eigenvalues are used to calculate the single-particle Green's function in imaginary time,
\begin{equation}
\label{green_spacetime}
  G (\br, \br', i \tau) = \begin{cases}
    \;\;\;\, i \sum\limits_i^{\tmop{occ}} \psi_i (\br) \psi_i^* (\br') e^{
    -|(\varepsilon_i^\text{DFT}-\varepsilon_\text{F}) \tau|}, & \tau < 0\;,\\[1em]
    - i \sum\limits_a^{\tmop{empty}}\hspace{-0.4em} \psi_a^* (\br) \psi_a (\br') e^{
    -|(\varepsilon_a^\text{DFT}-\varepsilon_\text{F}) \tau|}, & \tau > 0\;,
  \end{cases} 
\end{equation}
where the sum over the index $i$ runs over all occupied KS orbitals and the sum over the index $a$ over all virtual, i.e., empty KS orbitals. $\varepsilon_\text{F}$ is the Fermi level.
The irreducible polarizability follows,
\begin{equation}
\chi (\br, \br', i \tau) = - iG (\br,
   \br', i \tau) G (\br, \br', - i \tau)\;,
   \label{e3}
\end{equation}
which is then  transformed to imaginary frequency,
\begin{align}
\chi (\br, \br', i\omega)
  = i \int\limits_{-\infty}^\infty 
  e^{-i\omega\tau}\;
  \chi (\br, \br',i\tau)\;\mathd \tau\,.
\end{align}
This transform can be  understood as Laplace transform followed by analytic continuation to the imaginary axis and effectively takes the form of a Fourier transform~\cite{rieger1999gw}.
Next, the dielectric function $\epsilon$ can be calculated in imaginary frequency
from the irreducible polarizability,
\begin{equation} 
\epsilon (\br, \br', i \omega) = \delta (\br -
   \br') - \int \mathd \br'' v (\br,
   \br'') \,\chi (\br'', \br', i \omega)\;, \label{eps}
\end{equation}
using the Dirac delta function~$\delta(\br)$ and the Coulomb interaction $v (\br, \br')\eqt 1/|\br\mt\br'|$. The screened
Coulomb interaction can be computed from the inverse dielectric function,
\begin{equation} 
W (\br, \br', i \omega) = \int \mathd \br''\,
   \epsilon^{- 1} (\br, \br'', i \omega) \,v
   (\br'', \br')\;. \label{W}
\end{equation}
It is convenient in $GW$ implementations to split the screened interaction~$W$ into the bare Coulomb interaction~$v$ and the correlation part~$W^\text{c}$,
\begin{align}
   W^\text{c} (\br, \br', i \omega)  = W (\br, \br', i \omega) - v(\br, \br')\,.
\end{align}
In the space-time method, $W^\text{c}$ is required in imaginary time,
\begin{align}
  W^\text{c} (\br, \br', i\tau)
  = \frac{i}{2\pi} \int\limits_{-\infty}^\infty 
  e^{i\omega\tau}\;
   W^\text{c} (\br, \br',i\omega)\;\mathd \omega\,,
\end{align}
and the correlation self-energy is given as product of the  Green's function and the screened Coulomb interaction,
\begin{equation}
\label{sigma-ls}
 \Sigma^\text{c} (\br, \br', i \tau) = i\,G (\br,
   \br', i \tau) W^\text{c} (\br, \br', i \tau)\;.
\end{equation}
The self-energy is then transformed to imaginary frequency, 
\begin{align}
  \Sigma^\text{c} (\br, \br', i\omega)
  = i \int\limits_{-\infty}^\infty 
  e^{-i\omega\tau}\;
   \Sigma^\text{c} (\br, \br',i\tau)\;\mathd \tau\,,\label{Sigmactime}
\end{align}
and we calculate its $(n,n)$-diagonal element,
\begin{align}
    \Sigma_n^\text{c} (i\omega) &= \langle \psi_n| \Sigma^\text{c} (i\omega)|\psi_n\rangle \nonumber
    \\&= \int \mathd \br \,\mathd \br' \,\psi_n^*(\br)\, \Sigma^\text{c} (\br, \br', i\omega)\,\psi_n(\br')\,.
\end{align}
The self-energy is then analytically continued to real frequency~\cite{Golze2019,Leucke2025}.

Focusing on the $G_0W_0$ method already introduced before, we use KS orbitals to  approximate the QP wavefunctions and we compute $G$ and $W$ only once using KS orbitals and KS eigenvalues from Eqs.~\eqref{green_spacetime}\,--\,\eqref{W}.
The QP energies~$\varepsilon_n^{G_0 W_0}$ can finally be calculated by solving the QP equation,
\begin{equation}
\label{energy-it}
\varepsilon_n^{G_0 W_0} = \varepsilon_n^{\tmop{DFT}} + \tmop{Re} \Sigma_n^\text{c}
   (\varepsilon_n^{G_0 W_0}) + \Sigma_n^\text{x} - v_n^{\tmop{xc}}\,,
\end{equation}
where $ \Sigma_n^\text{x}$ and $v_n^{\tmop{xc}}$ are the $n,n$-diagonal elements of the exchange self-energy and the exchange-correlation potential.


\section{KS-DFT with periodic boundary conditions and Gaussian basis functions}
\label{sec3}
We use  KS-DFT with periodic boundary conditions~\cite{Delley2000,Martin2004,Blum2009}, i.e., $h_0 (\br) $ and $ v^\text{xc} (\br)$ from Eq.~\eqref{e1} are lattice periodic,
\begin{align}
 h_0 (\br+\bR) = h_0(\br)   \;,
 \hspace{2em}
 v^\text{xc} (\br+\bR) = v^\text{xc} (\br)\,,
\end{align}
for every lattice vector 
\begin{align}
    \bR\eqt \sum_{j=1}^{d} n_j\,\mathbf{a}_j\,,
\end{align} 
where $d$ is the dimension, $n_j$ are integers and $\mathbf{a}_j$ the primitive vectors of the lattice.
Bloch's theorem~\cite{Bloch1929} states that the solutions~$\psi_{n\bk}(\br)$ of the Kohn-Sham equations 
\begin{equation}
[h_0 (\br) + v^\text{xc} (\br)] \psi_{n\bk}  (\br) =
   \varepsilon_{n\bk} ^{\tmop{DFT}} \psi_{n\bk} (\br) \label{e14}
\end{equation}
with lattice-periodic~$h_0(\br)$ and $v^\text{xc} (\br)$ are Bloch functions
\begin{align}
   \psi_{n\bk} (\br) = e^{i\bk\cdot\br}u_{n\bk}(\br) \label{e16}
\end{align}
where $\bk$ is the crystal momentum in the first Brillouin zone (BZ) and $u_{n\bk}(\br) $ is a lattice-periodic function, $u_{n\bk}(\br)\eqt u_{n\bk}(\br\pt\bR)  $. 
The eigenvalues~$\varepsilon_{n\mathbf{k}}^\text{DFT}$ of band~$n$ and crystal momentum~$\mathbf{k}$  of the first Brillouin zone are the DFT bandstructure.
The requirement~\eqref{e16} on Bloch functions~$\psi_{n\mathbf{k}}(\br)$ can be fulfilled by the basis expansion
\begin{align}
\psi_{n\mathbf{k}}(\br) = \sum_\mu C_{ \mu n}(\mathbf{k})\sum_{\bR}\eikR\phi^\bR_\mu(\br)\,,\label{kpointsbf}
\end{align}
where~$C_{\mu n}(\mathbf{k})$  are molecular orbital (MO) coefficients and $\phi_\mu^\bR(\br)$ are atom-centered Gaussian-type basis functions  centered on an atom in the cell with lattice vector~$\bR$. 
For computing the MO coefficients $C_{\mu n}(\mathbf{k})$, one inserts Eq.~\eqref{kpointsbf} into Eq.~\eqref{e14}, multiplies with an atom-centered Gaussian function~$\phi_\nu^\bo(\br)$ in the unit cell~$\bo$, and integrates over  the whole real space, which gives
\begin{align}
\sum_\nu  \hks \, C_{ \nu n}(\mathbf{k})
=
\sum_\nu  S_{\mu\nu}(\bk) \, C_{ \nu n}(\mathbf{k})
\,  \varepsilon_{n\bk} ^{\tmop{DFT}}\,,\label{e18a}
\end{align}
with the Kohn-Sham matrix 
\begin{align}
\begin{split}
& \hks  = \sum_\bR \,\eikR\;
h_{\mu\nu}^\bR\;,\label{e18}
\\[0.3em]
&h_{\mu\nu}^\bR =  \int d\br\;\phi_\mu^\bo(\br) \left[h_0(\br)+v^\text{xc}(\br)\right]\phi_\nu^\bR(\br)
\,,
\end{split}
\end{align}
and the overlap matrix 
\begin{align}
 &S_{\mu\nu} (\bk)  = \sum_\bR \,\eikR 
\; S_{\mu\nu}^\bR \;,
\hspace{1em}
S_{\mu\nu}^\bR =  
 \int d\br\;\phi_\mu^\bo(\br) \,\phi_\nu^\bR(\br)\,.
 \label{e19}
\end{align}
Note that the sums over lattice vectors~$\bR$ in Eqs.~\eqref{e18} and~\eqref{e19} can be restricted to $\bR$ with small~$|\bR|$ because the atom-centered Gaussian function~$\phi_\mu^\bR(\br)\equivt \phi_\mu^\bo(\br\mt\bR) $ quickly decays for large~$|\br\mt\bR|$.

The Kohn-Sham matrix~$\hks$ depends on the electron density~$n(\br)$,
\begin{align}
       n(\br) &= \sum_i^\text{occ} \intbz\; |\psi_{i\bk}(\br)|^2 \,,\label{e21a}
\end{align}
where we integrate over the crystal momentum~$\bk$ in the BZ and $\Omega_\text{BZ}$ denotes the volume of the BZ. 
To obtain an efficient algorithm for computing~$n(\br)$, we use Eq.~\eqref{kpointsbf} to arrive at
\begin{align}
   n(\br)  =
   \sum_{\mu\nu} \sum_{\bR_1\bR_2} D_{\mu\nu}^{\bR_2-\bR_1} \;\phi_\mu^{\bR_1}(\br)\,\phi_\nu^{\bR_2}(\br)
    \label{e22a}
\end{align}
using the density matrix
\begin{align}
&D_{\mu\nu}^{\bR} = \intbz\; \emikR \;D_{\mu\nu}(\bk)\,, \label{e23}
  \\[0.5em]
& D_{\mu\nu}(\bk) =  \sum_n^\text{occ} C_{\mu n} (\bk)\,C_{\nu n}^*(\bk)\,.\label{e24a}
\end{align}
The integration over the BZ in Eq.~\eqref{e23} is executed in every self-consistent field cycle of the KS-DFT calculation using a discrete~$N_1\timest N_2\timest N_3$ Monkhorst-Pack $k$-point mesh~$\{\bk_\ell\}$~\cite{Monkhorst1976} for two-dimensional periodicity:
\begin{align}
  &D_{\mu\nu}^{\bR} \simeq \frac{1}{N_1N_2\hspace{0.05em}N_3} \sum_{\bk_\ell}^\text{BZ} e^{-i\bk_\ell\cdot\bR} \;\sum_n^\text{occ} C_{\mu n}(\bk_\ell)\,C_{\nu n}^*(\bk_\ell)\,.\label{e24} 
\end{align}
For a periodic direction~$j$, we choose~$N_j$ as even  integers, which leads to a $k$-mesh that  excludes the $\Gamma$-point,
\begin{align}
  \bk_\ell   = \sum_{j=1}^{d}  \frac{ \ell_j}{2N_j}\,\bb_j\;,\hspace{1em}
  \label{kmesh}
\end{align}
where we define $\ell\eqt\{\ell_j\}_{j=1}^{d}$ and each $\ell_j$ takes one of the following odd integers
\begin{align}
\ell_j\in \{\pm\,1,\, \pm\,3, \;\ldots\;,\, \pm\, (N_j-1)\}
\,.\label{ell}
\end{align}
$\bb_j$ are the primitive translation vectors of the reciprocal lattice that fulfill $\ba_{j_1}\cdott \bb_{j_2}\eqt 2\pi\delta_{j_1j_2}$.

Note that the density matrix~$D_{\mu\nu}^{\bR} $ computed from Eq.~\eqref{e24} features an erroneous   periodicity in a superlattice with primitive translation vectors~$\{\bT_j\}_{j=1}^d$,
\begin{align}
    D_{\mu\nu}^{\bR+\bT} = D_{\mu\nu}^{\bR}
        \;,    \hspace{1.3em}
    \bT = \sum_{j=1}^d  t_j  \bT_j
    \;,    \hspace{1.3em}
    \bT_j = 2 N_j\, \ba_j\;,
\end{align}
where $\{t_j\}_{j=1}^d$ are integers.
To avoid issues in a practical KS-DFT calculation, we  restrict the lattice vectors~$\bR$ in Eq.~\eqref{e24} to a single supercell (SC) of the $\bT$-superlattice, 
\begin{align}
    \bR\in \text{SC}
    \hspace{0.4em}
    \Leftrightarrow
    \hspace{0.4em}
    \bR = \sum_{j=1}^d n_j\,\ba_j
    \;,\hspace{0.3em}
    n_j\in \{0\,,\,\pm\,1\,,\,\ldots\,,\,\pm\,N_j\}\,.\label{e28}
\end{align}
\reviewnew{This restricts the lattice summation  in Eq.~\eqref{e22a} to ${\bR_1}{-}{\bR_2}\intext\text{SC}$.}
In the limit of a dense $k$-point mesh, i.e., large $N_1,N_2,N_3$, the SC~\eqref{e28} is large and the lattice summation~\eqref{e22a} converges, because the overlap~$\phi_\mu^{\bR_1}(\br)\,\phi_\nu^{\bR_2}(\br)$ quickly decays for large $|\bR_1\mt\bR_2|$.

\ASH{Starting from the eigenvalues $\varepsilon_{n\bk} ^{\tmop{DFT}}$ from Eq.~\eqref{e18a}, we  add  spin-orbit coupling (SOC) $V^{\tmop{SOC}}$ from Hartwigsen-Goedecker-Hutter (HGH) pseudopotentials~\cite{Hartwigsen1998,Krack2005,Vogt2025}, see details in Appendix~\ref{sec:SOC}, to obtain the spin-orbit perturbed Hamiltonian:
\begin{align}
h^{\tmop{DFT+SOC}}_{n\sigma,\,n'\sigma'}\hspace{-0.1em}(\bk)
=
\delta_{nn'}\,\delta_{\sigma\sigma'}\,\varepsilon_{n\mathbf{k}}^{\tmop{DFT}}
+ 
   V_{nn',\sigma\sigma'~}^{\tmop{SOC}}\hspace{-0.2em}(\bk)\,.
\end{align}
We diagonalize the Hamiltonian with SOC to obtain the DFT band structure with SOC:
\begin{align}
 \sum_{n'\sigma'} 
 h^{\tmop{DFT+SOC}}_{n\sigma,\,n'\sigma'}\hspace{-0.1em}(\bk)\;
 C_{n'\sigma'}^{(j)}(\bk)
 =
 \varepsilon^{\tmop{DFT+SOC}}_{j\bk} C_{n \sigma }^{(j)}(\bk)\,.\label{e31}
\end{align}
}

\section{\textit{GW} space-time method with Gaussian basis functions and periodic boundary conditions }\label{sec:IV}

In this section, we reformulate the $GW$ space-time method shown in Sec.~\ref{sec2} in an atomic-orbital basis with periodic boundary conditions.
The starting point is a DFT calculation with periodic boundary conditions in an atomic-orbital basis (Sec.~\ref{sec3}).
In the main text, we only present the working equations that are implemented in the algorithm.
We give a detailed derivation of these  equations in the Appendix.

As shown in Appendix~\ref{sec:derivationchi}, the Green's function in imaginary time is given by~\cite{liu2016}
\begin{align}
\begin{split}
G_{\mu \nu}(i\tau,\mathbf{k}) =& \;
\theta(\tau) \sum_a^\text{empty} C_{ \mu a}(\mathbf{k})C_{ \nu a}^*(\mathbf{k})e^{-(\varepsilon_{a\mathbf{k}}-\varepsilon_\text{F})\tau}
\\
& - \theta(-\tau) \sum_i^\text{occ} C_{ \mu i}(\mathbf{k})C_{ \nu i}^*(\mathbf{k})e^{-(\varepsilon_{i\mathbf{k}}-\varepsilon_\text{F})\tau}\,.
\end{split}
\label{e29}
\end{align}
Note that Eq.~\eqref{e29} is the analogue to Eq.~\eqref{e24a} for the density matrix.
As further shown  in Appendix~\ref{sec:derivationchi}, the Green's function is transformed to real space via an integration over the BZ,
\begin{align}
 G^\bR_{\mu \nu}(i\tau) &= \intbz\;\emikR \,G_{\mu \nu}(i\tau,\mathbf{k}) \\
&\reviewnew{\simeq \frac{1}{N_1N_2N_3}\sum_{\kl}^\text{BZ}\emiklR \,G_{\mu \nu}(i\tau,\kl)}
\,.\label{e21}
\end{align}
\reviewnew{The $\kl$-mesh used in Eq.~\eqref{e21} is the odd Monkhorst-Pack mesh from the DFT calculation used for the density matrix, Eqs.~\eqref{e24}\,--\,\eqref{ell}.}

\reviewnew{For the computation of the density response~$\chi$, we introduce the Resolution of Identity (RI)  \cite{Vahtras1993}, where products of Gaussian basis functions are expanded over an auxiliary RI basis set~$\{\varphi_P^\bP\}$:
\begin{align}
    \phi_\mu^{\bR}(\br)\,\phi_\nu^{\bT}(\br)
    \simeq
    \sum_{P\mathbf{P}} B_{P\bP}^{\mu\bR\nu\bT}\varphi_P^\bP(\br)\,,\label{eRI}
\end{align}
where the projection coefficients are defined as
\begin{align}
B_{P\bP}^{\mu\bR\nu\bT}
=\sum_{Q\bQ} 
(\mu\bR \,\nu\bT\,|\,Q\bQ)\,(M^{-1})_{PQ}^{\bP-\bQ}\,,\label{epRI}
\end{align}
with the three-centre integrals
\begin{align}
&(\mu\bR\,\nu\bT \,|\,P\bP ) = {\int} d\br\,d\br'\,
\phi_\mu^{\bR}(\mathbf{r})\,
\phi_\nu^{\bT}(\mathbf{r})\,
\Vt \,
\varphi_P^{\bP}(\mathbf{r}') \label{e5c}
\end{align}
that can be understood as projection of the product~$\phi_\mu^{\bR_1}
\phi_\nu^{\bR_2}$ onto the RI basis function $\varphi_P^\bR$ under a metric, which we have chosen to be the truncated Coulomb operator (RI-tCm)~\cite{Jung2005,wilhelm2021,Bussy2024},
\begin{align} 
&    \Vt=\left\{ 
    \begin{array}{cl}
        1/|\br-\br'| &\hspace{1em} \text{if } |\br-\br'|\le \rc \,, \\[0.5em]
        0 &\hspace{1em} \text{else\,.}
    \end{array}
    \right.  \label{e6c}
\end{align}
The inverse metric matrix $(M^{-1})_{PQ}^{\bP-\bQ}$ in Eq.~\eqref{epRI}  arises because of the non-orthogonality of the auxiliary RI basis set and is computed from
\begin{align}
M_{PQ}(\mathbf{k})  =&
\sum_\mathbf{R}\eikR \int  d\mathbf{r}\,d\mathbf{r}'\,
\varphi_P^\bo(\mathbf{r}) \, \Vt \,\varphi_Q^\mathbf{R}(\mathbf{r}')\,, \label{e11a}
\end{align}
a subsequent matrix inversion and transformation to real space, as shown in detail in Appendix~\ref{app:RI} which contains a derivation of Eq.~\eqref{eRI}/\eqref{epRI}.

%
The cutoff radius $\rc$ from Eq.~\eqref{e6c} is typically chosen in the order of a few {\AA}ngstroms~\cite{wilhelm2021,Bussy2024, Graml2024}. 
In the limit  $\rc\rightarrowtext0$, RI-tCm is equivalent to the RI with the overlap metric which suffers from a slow convergence with respect to the RI basis set size~\cite{Vahtras1993, Jung2005}.
In the limit of $\rc\rightarrowtext\infty$, RI-tCM is equivalent to the RI with the Coulomb metric where the convergence with the RI basis set size is fast. 
The drawback of the Coulomb metric is that tensor elements $(\mu\bR\,\nu\bT \,|\,P\bP ) $ from Eq.~\eqref{e5c} with the Coulomb operator only decay polynomially for $|\bR{-}\bP|\rightarrowtext\infty$ and  $|\bT{-}\bP|\rightarrowtext\infty$ which prohibits the application of RI with the Coulomb metric in this algorithm.
When truncating the Coulomb operator as in Eq.~\eqref{e6c} at finite $\rc$ and when using Gaussian basis functions,  the tensor~$(\mu\bR\,\nu\bT \,|\,P\bP ) $ decays like a Gaussian for 
$|\bR{-}\bP|\rightarrowtext\infty$ and  $|\bT{-}\bP|\rightarrowtext\infty$.
It has been shown that RI-tCm  converges quickly with the size of the auxiliary basis~$\{\varphi_P^\bR\}$.~\cite{Jung2005,wilhelm2021}
}
The density response~$\chi_{PQ}^\bR(i\tau)\eqt \braket{\varphi^\bo_P|\chi(i\tau)|\varphi^\bR_Q}$ in imaginary time in  the RI basis~$\{\varphi^\bR_P\}$ can be obtained as (see Appendix~\ref{sec:derivationchi} for a derivation and Appendix~\ref{sec:parallelchi} for parallel implementation)
\begin{align}
\chi_{PQ}^\bR(i\tau) =& 
 \sum_{\lambda\bR_1\nu\bR_2}
\bigg[\sum_{\mu } \sum_{\mathbf{S}_1}^\text{SC}
(\mu\bR_1{-}\mathbf{S}_1\,\nu\bR_2\,|\,P\bo) \;
G^{\mathbf{S}_1}_{ \lambda\mu}(-i\tau)\bigg]
\nonumber
\\
&
\hspace{1.0em}\times \bigg[\sum_{\sigma}\sum_{\mathbf{S}_2}^\text{SC}
(\lambda\bR_1\,\sigma\bR_2{-}\mathbf{S}_2\,|\,Q\bR)\;
G^{\mathbf{S}_2}_{\nu \sigma}(i\tau)\bigg]\,,
 \label{chiT}
\end{align}
where $\bR$ is a lattice vector inside the SC [Eq.~\eqref{e28}], $\bR_1,\bR_2,\bS_1,\bS_2$ are lattice vectors.  
\reviewnew{Note that we only sum over cells~$\bR_1$ and~$\bR_2$ with  small $|\bR_1|,|\bR_2|$ because of the sparsity of the three-centre integrals~\eqref{e5c}.}

Following the $GW$ space-time method,~\cite{rojas_1995,Graml2024} we transform the polarizability from real space and time to the Brillouin zone and frequency
\begin{align}
\chi_{PQ}(\mathbf{k},i\omega) =\sum_\bR^\text{SC} \int d\tau \, \cos(\omega\tau) \,\eikR\; \chi_{PQ}^\bR(i\tau)\,.\label{chitrafo}
\end{align}
\reviewnew{Note that as $\chi^\bR$ computed from Eq.~\eqref{chiT} decays exponentially for large $\bR$ in gapped systems, so one can perform the Fourier transformation~\eqref{chitrafo} to an arbitrary $\bk$.}
We execute the imaginary-time integration numerically using minimax grids~\cite{liu2016,Azizi2023,Azizi2024}.

As next step, we calculate the dielectric function~\cite{Graml2024}
\begin{align}
{\boldsymbol{\epsilon}}(\mathbf{k},i\omega) =  \mathbf{Id}- \mathbf{V}^{0.5}(\mathbf{k}) \bM^{-1}(\mathbf{k})\boldsymbol{\chi}(\mathbf{k},i\omega) \bM^{-1}(\mathbf{k})\mathbf{V}^{0.5}(\mathbf{k})
\label{e22}
\end{align}
where $\mathbf{Id}$ is the identity matrix and the truncated Coulomb matrix~$\bM(\bk)$ \reviewnew{defined in Eq.~\eqref{e11a}} appears due to the  RI-tCm. 
We  use Tikhonov regularization~\cite{Goncharsky1995} for the RI expansion to prevent linear dependencies of fit coefficients, as discussed in the supporting information of Ref.~\cite{Graml2024}. 
The regularization leads to the modified matrix inversion, 
\begin{align}
     \mathbf{M}^{-1}(\bk) = \big(\mathbf{M}(\bk)\pt\alpha\mathbf{Id}\big)^{-1}\,,\label{e12}
\end{align}
where $\alpha$  is the regularization parameter. 
%

$\mathbf{V}^{0.5}(\mathbf{k})$  in Eq.~\eqref{e22} is the matrix square root~of the bare Coulomb interaction~$\mathbf{V}(\mathbf{k})$~\cite{Grundei2017, zhu2021all, Garcia2025,spadetto2025}
\begin{align}
V_{PQ}(\mathbf{k}) &= 
\sum_\bR e^{i\bk\cdot\bR}{\int} d\br\,d\br'\,\varphi_P^\bo(\br)\,\frac{ 1}{|\br- \br'|}\,\varphi_Q^\bR(\br')\,.\label{Vper}
\end{align}
Details on the lattice summation over~$\bR$ are given in Appendix~\ref{app:Vk}.
We obtain the correlation part of the screened interaction $W^\text{c}(i\omega) \eqt (\epsilon^{-1}(i\omega)\mt1)V  $ as
\begin{align}
\mathbf{W}^\text{c}(\mathbf{k},i\omega) =  \mathbf{V}^{0.5}(\mathbf{k})({\boldsymbol{\epsilon}}^{-1}(\mathbf{k},i\omega) -\mathbf{Id}) \mathbf{V}^{0.5}(\mathbf{k})
\label{Wk}
\end{align}
and  transform it to real space $W_{PQ}^{\text{c},\bR}\eqt \braket{\varphi^\bo_P|W^\text{c}|\varphi^\bR_Q}$,
\begin{align}
W_{PQ}^{\text{c},\bR} (i\omega) = \intbz\; \emikR \,W^\text{c}_{PQ}(\mathbf{k},i\omega)\,.\label{trafoWtoR}
\end{align}
Special care is required for the BZ integral as $W_{PQ}^\text{c}(\bk,i\omega)$ diverges at the $\Gamma$-point with $1/k$ for two-dimensional materials if $\varphi_P$ and $\varphi_Q$ are s-type basis functions~\cite{Wilhelm2017,Ren/etal:2021,zhu2021all}. 
%
%
We evaluate  $W_{PQ}^\text{c}(\mathbf{k},i\omega)$ using two Monkhorst-Pack meshes~\cite{Monkhorst1976}: 
$\{\bk_\ell\}$ has $4N_j$ $k$-points in periodic directions~$j$ 
  and~$\{\bq_\ell\}$ has $8N_j$ $k$-points in periodic directions~$j$.
The number of $k$-points in the~$\{\bk_\ell\}$ mesh and $\{\bq_\ell\}$ mesh is thus
\begin{align}
    N_k = \prod_{j=1}^d 4N_j\;,
    \hspace{2em}
    N_q = \prod_{j=1}^d 8N_j\;.
\end{align} 
We extrapolate the BZ integration~\eqref{trafoWtoR} with the inverse square root of the number of $k$-points~\cite{zhu2021all}.
Reformulating Eq.~\eqref{trafoWtoR} , the $k$-extrapolated screened Coulomb interaction becomes
\begin{align}
    W_{PQ}^{\text{c},\bR}(i\omega) 
&= 
 \sum_{\ell}  v_\ell\,e^{-i\mathbf{q}_\ell\cdot\bR} \,W_{PQ}^\text{c}(\mathbf{q}_\ell,i\omega) \nonumber
\\ &
 - \sum_{\ell}  w_\ell\,e^{-i\mathbf{k}_\ell\cdot\bR} \,W_{PQ}^\text{c}(\mathbf{k}_\ell,i\omega)\,,
 \label{e16a}
\end{align}
where the extrapolation is incorporated into the integration weights:
\begin{align}
    v_\ell =  
       \frac{1}{(1-\sqrt{N_k/N_q })\,N_q}\;\;,
       \hspace{1em}
    w_\ell =  
       \frac{1}{(\hspace{-0.2em}{\sqrt{N_q/N_k }}-1)\,N_k}\;.   
        \label{e17b}
\end{align}

We transform $ W_{PQ}^{\text{c},\bR}(i\omega) $ back to $k$-points, 
\begin{align}
  \hat{W}^\text{c}_{PQ}(\bk,i\omega) =   
  \sum_\mathbf{R}^\text{SC}\eikR\;
   W_{PQ}^{\text{c},\bR}(i\omega) \,,
\end{align}
 where we only sum over the cells $\bR\intext\text{SC}$, see Eq.~\eqref{e28}, to prevent for the divergence of $\hat{W}^\text{c}(\bk,i\omega)$ at  $\bk\eqt0$.
We incorporate the RI metric, 
\begin{align}
      \tilde{\mathbf{W}}^\text{c}(\bk,i\omega) 
      =
      \bM^{-1}(\bk)\, \hat{\mathbf{W}}^\text{c}(\bk,i\omega)\,\bM^{-1}(\bk)\,,
\end{align}
and transform to real space using 
the $k$-point \reviewnew{mesh from DFT, Eqs.~\eqref{e24} and~\eqref{kmesh}}
\begin{align}
 \tilde{W}_{PQ}^{\text{c},\bR} (i\omega) &= \intbz\; \emikR \,\tilde{W}^\text{c}_{PQ}(\mathbf{k},i\omega) \\
&\reviewnew{\simeq \frac{1}{N_1N_2N_3}\sum_{\kl}^{\textrm{BZ}}\emiklR \,\tilde{W}^\text{c}_{PQ}(\kl,i\omega)\,.}\label{e49}
\end{align}

Following the $GW$ space-time method~\cite{rojas_1995}, we transform $W^\text{c}(i\omega)$ to imaginary time using minimax grids~\cite{liu2016,Azizi2023,Azizi2024}.
This completes the ingredients for the $GW$ self-energy $\Sigma^\text{c}(i\tau)\eqt iG(i\tau)W^\text{c}(i\tau) $. 
We calculate the self-energy~$\Sigma^{\text{c},\bR}_{\lambda\sigma}(i\tau)\eqt \braket{\phi_\lambda^\bo |\Sigma^\text{c}(i\tau)|\phi_\sigma^\bR}$  in real space (derivation in Appendix~\ref{sec:derivationsigma}), 
\begin{align}
\Sigma^{\text{c},\bR}_{\lambda\sigma}(i\tau) &=
i
\sum_{P\nu} \sum_{\bR_1 }^\text{SC}
\sum_{ \mathbf{S}_1}^\text{SC}
\left[
\sum_\mu
\sum_{ \mathbf{S}_2}^\text{SC}(\lambda\bo\;\mu\mathbf{S}_1{-}\mathbf{S}_2 \,|\,P\bR_1)
\; G^{\mathbf{S}_2}_{\mu\nu}(i\tau)\right] \nonumber
\\[0.2em]
&\hspace{1em}\times
\left[\sum_Q\sum_{\bR_2 }^\text{SC}         
( \sigma\bR\;\nu\mathbf{S}_1 \,|\,Q\bR_1{-}\bR_2)
\;\tilde  W_{QP}^{\text{c},\bR_2}(i\tau)\right] 
\label{SigmaR}
\end{align}
and in the Kohn-Sham basis~$\psi_{n\bk}(\br)$,
\begin{align}
&\Sigma_{\lambda\sigma}^\text{c}(\bk,i\tau) =
 \sum_\bR^\text{SC} \eikR\;
\Sigma_{\lambda\sigma}^{\text{c},\bR}(i\tau) \,,\label{e48}
\\[0.3em]
&\Sigma_{n\bk}^\text{c}(i\tau) = \sum_{\lambda\sigma}
C_{\lambda n}^*(\bk) \, \Sigma^\text{c}_{\lambda\sigma}(\bk,i\tau)\,C_{\sigma n}(\bk)\;.
\label{e17}
\end{align}

\ASH{A major advantage of the present $GW$ algorithm is that  the  self-energy is computed in  real space, Eq.~\eqref{SigmaR}, involving only a modest number of neighbor cells $\bR_1,\bR_2,\bS_1,\bS_2$ to be included.
Instead, when using $k$-point sampling instead of lattice summation to compute the self-energy, special care needs to be taken due to the diverging $W(\bk)$ at the $\Gamma$-point, which requires special correction schemes~\cite{FREYSOLDT20071}.
In the present algorithm, the divergence of $W(\bk)$ at the $\Gamma$-point is taken into account  via BZ integration of $W(\bk)$ in Eq.~\eqref{trafoWtoR} with suitable $k$-point grids, Eq.~\eqref{e16a}.
}

We calculate the exchange self-energy similarly to previous work~\cite{Irmler2018,Bussy2024},
\begin{align}
\Sigma^{\text{x},\bR}_{\lambda\sigma} &=
i
\sum_{P\nu} \sum_{\bR_1 }^\text{SC}
\sum_{ \mathbf{S}_1}^\text{SC}
\left[
\sum_\mu
\sum_{ \mathbf{S}_2}^\text{SC}(\lambda\bo\;\mu\mathbf{S}_1{-}\mathbf{S}_2 \,|\,P\bR_1)
\; D^{\mathbf{S}_2}_{\mu\nu}\right] \nonumber
\\[0.2em]
&\hspace{1em}\times
\left[\sum_Q\sum_{\bR_2 }^\text{SC}         
( \sigma\bR\;\nu\mathbf{S}_1 \,|\,Q\bR_1{-}\bR_2)
\; \tilde V_{QP}^{\text{tr},\bR_2} \right]
\,,
\label{SigmaxR}
\end{align}
where $D^{\mathbf{S}}_{\mu\nu}$ is the density matrix~\eqref{e23} from KS-DFT and 
\begin{align}
&\tilde {\mathbf{V}}^{\text{tr},\bR} 
= \intbz \;\emikR\;
  \bM^{-1}(\bk)\, \mathbf{V}^\text{tr}(\bk)\,\bM^{-1}(\bk)\,,
\\[0.5em]
  & V_{PQ}^\text{tr}(\bk) = 
   \sum_\bR \eikR
    \int  d\mathbf{r}\,d\mathbf{r}'\,
\varphi_P^\bo(\mathbf{r}) \, \Vtbvk \,\varphi_Q^\mathbf{R}(\mathbf{r}')
\end{align}
is the truncated Coulomb matrix with truncation radius $\rbvk\eqt \min_j N_j|\ba_j|/2$, i.e., half of the shortest primitive translation vector of the supercell SC~\eqref{e28}~\cite{Spencer2008,Irmler2018}. 
Note that we restrict all lattice vector differences appearing in Eq.~\eqref{SigmaR} and~\eqref{SigmaxR} to the SC~\eqref{e28}, i.e., $\bS_1{-}\bS_2,\bR_1{-}\bR_2\intext\text{SC}$.
We obtain the quasiparticle energies~$\varepsilon_{n\mathbf{k}}^{G_0W_0}$ by replacing the DFT exchange-correlation contribution~$v^\text{xc}_{n\mathbf{k}}$ with the self-energy, 
\begin{align}
\varepsilon_{n\mathbf{k}}^{G_0W_0} = \varepsilon_{n\mathbf{k}}^\text{DFT}+ \Sigma^\text{x}_{n\mathbf{k}} + \text{Re}\,\Sigma^\text{c}_{n\mathbf{k}}(\varepsilon_{n\mathbf{k}}^{G_0W_0}) -v^\text{xc}_{n\mathbf{k}}  \,.\label{qpeq}
\end{align}

\ASH{Identically to the DFT band structure~\eqref{e31}, we add the spin-orbit potential $V^{\tmop{SOC}}$ to the eigenenergies to obtain the spin-orbit corrected Hamiltonian (the details are given in Appendix~\ref{sec:SOC}):
\begin{align}
h^\gwsoc_{n\sigma,\,n'\sigma'}\hspace{-0.1em}(\bk)
=
\delta_{nn'}\,\delta_{\sigma\sigma'}\,\varepsilon_{n\mathbf{k}}^{G_0W_0}
+ 
   V_{nn',\sigma\sigma'~}^{\tmop{SOC}}\hspace{-0.2em}(\bk)\,.
\end{align}
Diagonalization  leads to the  $GW$ band structure with SOC:
\begin{align}
 \sum_{n'\sigma'} 
 h^\gwsoc_{n\sigma,\,n'\sigma'}\hspace{-0.1em}(\bk)\;
 C_{n'\sigma'}^{(j)}(\bk)
 =
 \varepsilon^\gwsoc_{j\bk} C_{n \sigma }^{(j)}(\bk)\,.
 \label{e58}
\end{align}
}

\RP{The primary difference between our algorithm and the $\Gamma$-point-only $GW$ implementation using Gaussian basis functions reported in Ref.~\cite{Graml2024} lies in the treatment of periodic boundary conditions. 
In our approach, both the irreducible density response function $\chi$ and the self-energy $\Sigma$ are computed via explicit lattice summations [Eqs.~\eqref{chiT} and~\eqref{SigmaR}] over all unit cells. 
In contrast, Ref.~\cite{Graml2024} restricts the lattice sum to the nearest-neighbor cell, effectively circumventing the need for a full lattice summation. 
This approach is only exact in the limit of large unit cells, and the presented $GW$ algorithm enables the treatment of small unit cells. 

}

\section{Computational Details}\label{sec5}

\ASH{
\subsection{Overview}
There are several numerical approaches, also involving approximations, to compute $GW$ band structures. 
Commonly employed approximations are the use of pseudopotentials to exclude core electrons from the computation, and the use of plasmon-pole models to simplify the frequency dependence in $GW$. 
In this work, we employ pseudopotentials, details given in the following, but we avoid plasmon-pole models by treating the full frequency (and time) dependence. 
The evaluation of intermediate quantities in the $GW$ method, which depend on the real-space coordinates~$\br$ and~$\br'$, requires the use of a basis set. 
Real-space grids, while conceptually straightforward, typically involve a large number of grid points and are therefore computationally less efficient. Common alternatives include plane-wave and atomic-orbital basis sets.
When using identical pseudopotentials—or when the influence of the pseudopotential is negligible—different basis sets yield the same $GW$ band structure, provided the basis is sufficiently large to ensure convergence. In practice, agreement between different $GW$ implementations is typically within 0.1~eV when the same pseudopotentials are used across codes~\cite{Rohlfing1993,rangel2020reproducibility}.
A comprehensive benchmark study comparing $GW$ results across different basis sets (e.g., plane-waves vs.\ atom-centered orbitals) for a large and diverse set of materials—on the order of 100 solids—with agreement at the 10~meV level or better has yet to be conducted, \reviewnew{and the very strong sensitivity of the band gaps with respect to parameters such as the lattice constant~\cite{Zollner2019} makes comparison with already available theoretical literature fairly difficult.} 
\RP{To assess the numerical precision of our $GW$ implementation (Sec.~\ref{sec:IV}), we carry out illustrative  tests on four selected reference materials, namely monolayer MoS$_\text{2}$, 
MoS$\text{e}_\text{2}$, 
WS$_\text{2}$ and WS$\text{e}_\text{2}$.} We use two well-established $GW$ codes, BerkeleyGW~\cite{Deslippe2012} and VASP~\cite{Klimes2014}, to compute reference band gaps and band structures for comparison.
}

\subsection{Atomic geometries of MoS$_\text{2}$, 
MoS$\text{e}_\text{2}$, 
WS$_\text{2}$, 
WS$\text{e}_\text{2}$}
For our benchmark calculations, we employ  monolayer transition metal dichalcogenides   MoS$_\text{2}$, 
MoS$\text{e}_\text{2}$, 
WS$_\text{2}$,  and 
WS$\text{e}_\text{2}$.
\ASH{These materials are non-magnetic and stable, and they attract widespread interest thanks to a rare combination of properties: they are atomically thin, have a direct bandgap and strong spin–orbit coupling which make them ideal for both fundamental studies and emerging applications in electronics, spintronics, optoelectronics and energy harvesting~\cite{Manzeli2017}.}
We take atomic geometries and lattice parameters for these materials from the C2DB database ~\cite{Gjerding2021}.

\subsection{\textit{GW} space-time calculations (CP2K)} \label{subsec:compparamcp2k}
We have implemented the low-scaling $GW$ space-time algorithm presented in this work  in the CP2K software package~\cite{Kuehne2020,Kuehne2025,cp2k}.
CP2K employs a Gaussian basis set for representing KS orbitals [Eq.~\eqref{kpointsbf}] and a plane-wave basis set for the electron density to evaluate the Hartree potential via Ewald summation~\cite{Kuehne2020,Kuehne2025}.
We use Gaussian dual-space pseudopotentials~\cite{Hartwigsen1998}.
In the DFT calculation, we employ the PBE exchange-correlation functional~\cite{Perdew1996}.
\ASH{The plane-wave cutoff for the electron density  is set to $500$ Ry. This value was converged beforehand on the DFT-level results, as the plane-wave grid is not used for the $GW$ part of the calculations.}
In the $GW$ algorithm, we employ a minimax time-frequency grid~\cite{liu2016,Azizi2023,Azizi2024}. 
We compute two- and three-centre integrals over Gaussians using analytical schemes~\cite{Golze2017,libint}.  
The self-energy is analytically continued from imaginary frequency to the real frequency using a Pad\'{e} model~\cite{Vidberg1977,gw100,Leucke2025} with 16 parameters.
Unless otherwise noted, we employ a cutoff radius $r_c \eqt 7$\,\AA, an RI regularization $\alpha\eqt10^{-2}$ and a box height for the 2D materials of 15\,{\AA} (for computing the Fourier transform in the Hartree potential).
As already stated, the extrapolation of the $k$-integration of $W$ [Eq.~\eqref{e17b}] relies on two $k$-meshes, one 4 times denser than the equivalent DFT $k$-mesh along each direction and one 8 times denser (e.g. for a $32\timest32$ DFT $k$-mesh, the corresponding coarse W $k$-mesh is $128\timest128$ and the dense $k$-mesh is $256\timest256$).
The remaining computational parameters include the number of minimax time and frequency points~$N_{\tau/\omega}$~\cite{liu2016,Azizi2023,Azizi2024}, the DFT $k$-mesh [Eqs.~\eqref{e21a}, \eqref{e23}, \eqref{e29}; the DFT $k$-mesh also defines the SC supercell~\eqref{e28}], and the filter threshold for sparse matrix-tensor operations in Eqs.~\eqref{chiT}, \eqref{SigmaR} and~\eqref{SigmaxR}.
\ASH{In this paper, we will use two sets of parameters: \textit{tight} settings ($N_{\tau/\omega}\eqt30$,  DFT $k$-mesh: $32\timest32$, filter:~$10^{-11}$) which correspond to a reference set of parameters that we have defined through extensive testing in order to yield well-converged $GW$ band gaps, and a set of \textit{light} settings ($N_{\tau/\omega}\eqt 10$, DFT $k$-mesh: $24\timest24$, filter:~$10^{-6}$) that have sufficiently reduced memory and computational costs in order to be used for laptop calculations while still giving decently accurate results (quantitative values are given in Sec. \ref{sec:DFTband} and Sec. \ref{sec:GWband}). Note that the memory bottleneck comes from the parallelization strategy, as one can see in Appendix \ref{sec:parallelchi}, which implies the storage of all the three-centre integrals on each parallel group. In theory, one could also recompute these at each loop of the code, which would in practice prohibitively increase the computation time.} 
\ASH{We use the single-, double- and triple-zeta MOLOPT basis sets~\cite{vande2007} for expanding the KS orbitals, Eq.~\eqref{kpointsbf}.
These basis sets have been optimized for the total energy of the ground state so that they might exhibit a slow convergence behaviour for excited states, and therefore for band gaps at the $GW$ level.}
This motivated us to also use augmented single-, double- \ASH{and triple-}zeta Gaussian basis sets (aug-SZV-MOLOPT, aug-DZVP-MOLOPT \ASH{and aug-TZVP-MOLOPT}). 
We have created those by augmenting Gaussian SZV-MOLOPT, DZVP-MOLOPT and TZVP-MOLOPT bases~\cite{vande2007} of S, Se, Mo, W with an additional $s$, $p$, $d$ function, an additional $f$ function (for Mo, W and aug-DZVP-MOLOPT and aug-TZVP-MOLOPT of S, Se) and an additional $g$ function (all aug-TZVP-MOLOPT and aug-DZVP-MOLOPT of Mo, W).
We have optimized the contraction coefficients of the additional functions by optimizing the lowest five $GW$+Bethe-Salpeter excitation energies~\cite{Blase2020} of a molecular set~\cite{vande2007}.
We will report DFT and $GW$ with the original, i.e., non-augmented MOLOPT basis sets and the augmented MOLOPT basis sets. 

\ASH{
For the density response~\eqref{chiT}, dielectric function~\eqref{e22} and screened Coulomb interaction~\eqref{Wk}, an auxiliary Gaussian RI basis set is required. 
There is no general approach for constructing optimally sized RI basis sets, as their design is closely tied to the chosen AO basis. In the case of the widely used cc-pVNZ Dunning basis sets \cite{Dunning1989}, for example, the corresponding RI basis sets introduced in Ref. \cite{Weigend2002} provide a single, fixed RI basis for each AO basis. As a result, convergence with respect to the size of the RI basis is not commonly investigated. It is also possible to generate RI basis sets on the fly \cite{Stoychev2017}, although this usually leads to fairly large numbers of basis functions.} 
In our case, we have optimized the RI basis set~$\{\varphi_P\}$ by matching the RI-MP2 correlation energy~\cite{Weigend1998} of single atoms to the MP2 correlation energy. Unless stated otherwise, all reference calculations in this paper 
were carried out using an RI basis with a relative RI-MP2 correlation energy difference of $10^{-3}$ with respect to the corresponding MP2 correlation energy. 
We incorporate SOC via parameters from HGH Gaussian dual-space pseudopotentials~\cite{Hartwigsen1998,Krack2005,Vogt2025}, see Eqs.~\eqref{e31},~\eqref{e58} and  Appendix~\ref{sec:SOC} for details.
\ASH{We compute the SOC for states in a window of 40~eV, so 20~eV below the valence band maximum to 20~eV above the conduction band minimum, in order to avoid possible numerical instabilities with bands far from the gap. For the case of WSe$_2$, we chose a window of 20~eV for aug-DZVP-MOLOPT and aug-TZVP-MOLOPT, as discussed in Appendix~\ref{sec:SOC}.}
All inputs and outputs of the calculations are openly available, see data and code availability statement.

\subsection{Reference \textit{GW} calculations with a plane-wave-based algorithm (BerkeleyGW)}
\label{subsec:paramQEBGW}
{
\subparagraph*{\textsc{Quantum Espresso}.} We performed the \textsc{Quantum Espresso} (QE) \cite{Giannozzi2017} DFT calculations employing the PBE exchange-correlation functional \cite{Perdew1996}. Fully relativistic norm-conserving pseudopotentials were used, as provided by the PseudoDojo database \cite{Dojo2019}. A plane-wave energy cut-off of 100 Ry was applied, and the self-consistent charge density was converged on a $30 \times 30 \times 1$ $\mathbf{k}$-grid with a total energy convergence threshold of $10^{-9}$ Ry.

\subparagraph*{Berkeley{GW}.} Using the QE DFT energies and states, we performed for each material a one-shot $GW$ calculation ($G_0W_0$) using the BerkeleyGW package \cite{Hybertsen1986, Deslippe2012}. We considered the full spinor implementation of BerkeleyGW \cite{Barker2022}, which incorporates SOC non-perturbatively.
The dielectric matrix was computed with a dielectric cut-off of 25 Ry, considering a total of 3999 occupied and empty bands on a $12 \times 12 \times 1$ uniform $\mathbf{k}$-grid. 
For completeness, we computed the quasi-particle band-gap using the generalized plasmon-pole model of Hybertsen-Louie \cite{Hybertsen1986} (see Appendix \ref{app:GPP}) and the full-frequency evaluation of the self-energy.
In the explicit full frequency calculation, we used the contour-deformation method with the Adler-Wiser formula. We employed a frequency step and broadening of 0.25 eV, using 15 frequency points along the imaginary axis within the contour deformation approach. A low-frequency cutoff of 20.0~eV was set to restrict the real-axis integration range. To accelerate convergence in the vicinity of $|\mathbf{q}| \rightarrow 0$, a non-uniform neck subsampling approach was employed \cite{Qiu2017} and the spurious interactions between periodic replicas in the perpendicular direction to the surface were removed with a Coulomb interaction truncation scheme \cite{Ismail2006}. The full frequency dependence of the self-energy was evaluated with a frequency step of $0.2$~eV in the frequency range $[-2.0, 2.0]$~eV and centering the frequency grid around each mean-field quasi-particle energy. For the band structures, the quasi-particle energies computed on the coarse $\mathbf{k}$-grid were interpolated along high-symmetry $\mathbf{k}$-paths.
}

 \begin{figure*}[t!]
    \centering
\includegraphics[width=1.0\textwidth]{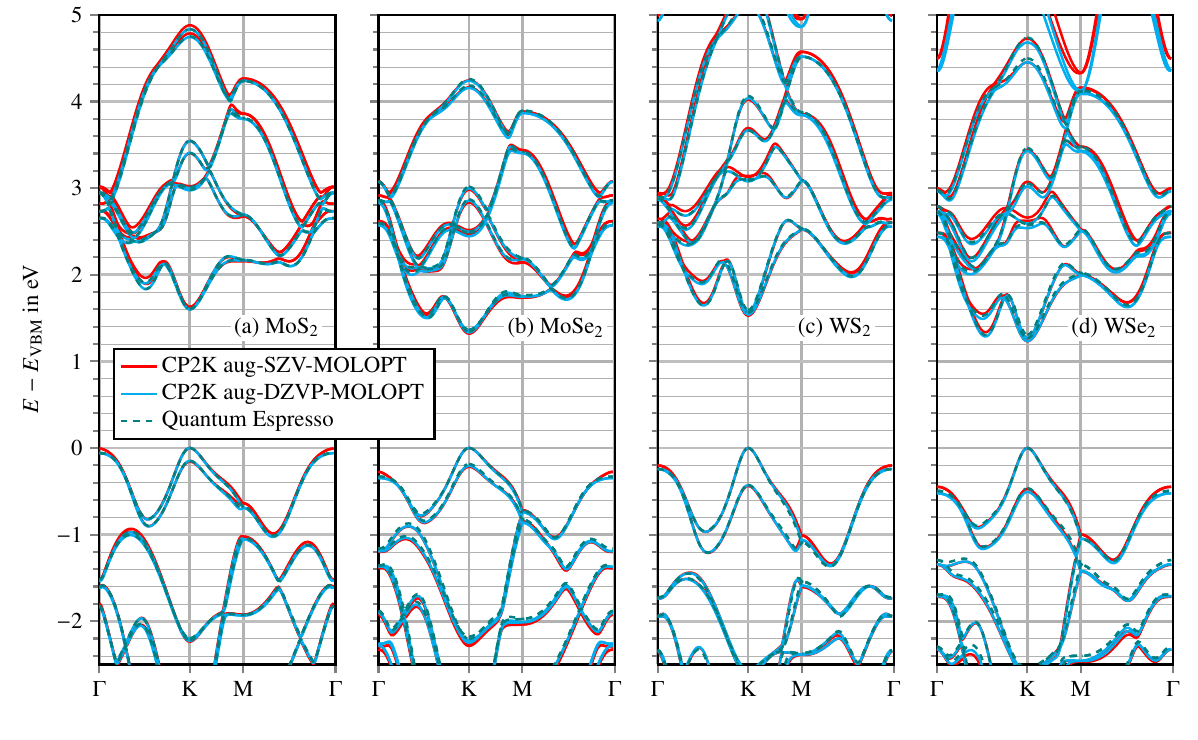}\vspace{-2em}
    \caption{PBE+SOC Bandstructures of monolayer MoS$_2$, MoSe$_2$, WS$_2$ and WSe$_2$, computed from Eq.~\eqref{e31} using Gaussian basis sets (CP2K) and a plane-wave basis (QE).  The computational details are given in Sec.~\ref{subsec:compparamcp2k} and \ref{subsec:paramQEBGW}.  The numerical values of the direct band gap at K are reported in Table~\ref{t1}.  }
    \label{fig7}
\end{figure*}

\subsection{Reference \textit{GW} calculations with the projector-augmented-wave scheme (VASP)}

To obtain the quasi-particle band gap, we carry out a single-shot $G_0W_0$ calculation using the Vienna Ab initio Simulation Package (\textsc{VASP}) \cite{vasp1, vasp2}. In this process, the initial KS wavefunctions and energy levels, derived from a preceding DFT calculation also performed with \textsc{VASP}, are used. Core electrons are treated using the $GW$ adaptation of the projector-augmented-wave (PAW) pseudopotentials \cite{PAW1, PAW2}. For all TMDCs, we apply PAWs constructed on the Perdew–Burke–Ernzerhof (PBE) functional \cite{Perdew1996} with an energy cutoff of 500 eV.
The considered valence electron configurations are $4s^24p^65s^14d^5$ for Mo, $3s^23p^4$ for S, $5p^66s^25d^4$ for W, $4s^24p^4$ for Se. 
A total of 384 bands, comprising both occupied and unoccupied states, are taken into account, along with a uniform $\mathbf{k}$-grid of $18 \times 18 \times 1$ for MoS$_2$ and WS$_2$, and of $15 \times 15 \times 1$ for MoSe$_2$ and WSe$_2$, ensuring smooth convergence in the vicinity of $|\mathbf{q}| \rightarrow 0$. Moreover, the dielectric tensor is broadened with a Lorentzian of 0.01~eV in all cases except for MoSe$_2$, where a broadening of 0.1~eV is applied. Finally, the self-energy is calculated using a full-frequency implementation with 100 points along the imaginary frequency grid, as provided in \textsc{VASP}.

\begin{table*}[]
  \fontsize{10}{12}\selectfont
 \caption{PBE bandgap, PBE+SOC bandgap [Eq.~\eqref{e31}],  $G_0W_0$@PBE+SOC bandgap [Eq.~\eqref{e58}] and K-point SOC splitting (in eV) of monolayer $\text{MoS}_\text{2}$, 
$\text{MoSe}_\text{2}$, 
WS$_\text{2}$ and
WS$\text{e}_\text{2}$ for various Gaussian MOLOPT basis sets~\cite{vande2007} with tight parameters (unless otherwise noted), and computed from the plane-wave codes \textsc{Quantum Espresso (QE)} ~\cite{Giannozzi2017}, {BerkeleyGW (BGW)}~~\cite{Deslippe2012} and \textsc{VASP} ~\cite{vasp1, vasp2}. The $GW$ calculations for the aug-TZVP-MOLOPT basis set are computationally intractable with our current implementation.
 }
 \vspace{0.5em}
{\small
\renewcommand{\arraystretch}{1.2}
    \begin{tabular}{lcc@{\hspace{-2pt}}c@{\hspace{-2pt}}c@{\hspace{15pt}}cc@{\hspace{-10pt}}c@{\hspace{-10pt}}c@{\hspace{15pt}}cc@{\hspace{-15pt}}c@{\hspace{-15pt}}c@{\hspace{15pt}}cc@{\hspace{-25pt}}c@{\hspace{-25pt}}c}
    \hline
      &&& \hspace{-25pt} PBE bandgap&&&&\hspace{-25pt}PBE+SOC bandgap&&&&\hspace{-35pt}K-point PBE+SOC splitting&&&&\hspace{-30pt}$G_0W_0$@PBE+SOC bandgap& \\
      & MoS$_2$ & MoSe$_2$ & WS$_2$ & WSe$_2$ & MoS$_2$ & MoSe$_2$ & WS$_2$ & WSe$_2$ & MoS$_2$ & MoSe$_2$ & WS$_2$ & WSe$_2$ & MoS$_2$ & MoSe$_2$ & WS$_2$ & WSe$_2$ 
    \\
    \hline
        SZV   & 1.620 & 1.492 & 1.858  &  1.576 & 1.611 & 1.373 & 1.590  &  1.231 & 0.168 & 0.210 & 0.474 & 0.581 &  2.37 & 1.86 & 2.46 & 1.91
            \\
    DZVP  & 1.697 & 1.441 & 1.833  &  1.557 & 1.614 & 1.325 & 1.560  &  1.254 & 0.155 & 0.199 & 0.433 & 0.476 & 2.42 & 1.98 & 2.44  &  2.00 \\
    TZVP   & 1.688 & 1.446 & 1.822  &  1.555 & 1.606 & 1.328 & 1.545  &  1.246 & 0.151 & 0.197 & 0.439 & 0.488 & 2.36 & 1.97 & 2.43  &  1.98  \\
    TZV2P   & 1.683 & 1.450 & 1.813  & 1.550 & 1.602 & 1.335 & 1.541  & 1.246 & 0.150 & 0.194 & 0.434 & 0.478 & 2.35 & 1.99 & 2.44  &  2.02 \\
    aug-SZV, light &  1.711 & 1.448 & 1.798 & 1.529 &  1.625 & 1.320 & 1.527 & 1.232 & 0.158 & 0.218 & 0.439 & 0.473 &  2.35 & 1.89 & 2.43 & 1.99 \\
    aug-SZV  &  1.711 & 1.448 & 1.798 & 1.529 &  1.625 & 1.320 & 1.527 & 1.232 &  0.158 & 0.218 & 0.439 & 0.473 & 2.34  & 1.92 & 2.38  &  1.96
            \\
    
 aug-DZVP   & 1.679 & 1.456 & 1.810  &  1.558 & 1.598 & 1.333 & 1.540  &  1.234 & 0.150 & 0.206 & 0.430  &  0.508 & 2.30  & 1.94& 2.34  &  1.93
            \\
aug-TZVP   & 1.682 & 1.451 & 1.812  &  1.552 & 1.599 & 1.327 & 1.538  & 1.237 & 0.151 & 0.197 & 0.440  &  0.503 & N/A  & N/A & N/A  &  N/A
            \\
\textsc{QE/BGW}  & 1.680 & 1.450 & 1.818 & 1.554 & 1.602 & 1.343 & 1.562 & 1.268 & 0.149 & 0.186 & 0.427 & 0.465 & 2.28 & 1.98 & 2.36 & 2.05
      \\
\textsc{VASP}  & 1.681 & 1.450 & 1.816 & 1.552 & 1.599 & 1.341 & 1.550 & 1.256 & 0.150 & 0.186 & 0.427 & 0.464 & 2.29 & 2.01 & 2.37 & 2.02 \\
 \hline
    \end{tabular}
}
\label{t1}
 \end{table*}

\section{DFT band structure of $\text{MoS}_\text{2}$, 
$\text{MoSe}_\text{2}$, 
WS$_\text{2}$, 
WS$\text{e}_\text{2}$}
\label{sec:DFTband}
\ASH{
We begin by analyzing the DFT band structures and band gaps of monolayer $\text{MoS}_2$, $\text{MoSe}_2$, $\text{WS}_2$, and $\text{WSe}_2$, with particular focus on the agreement between the three numerical approaches described in Sec.~\ref{sec5}, namely Gaussian basis sets versus plane waves, and different treatments of core electrons via pseudopotentials.
This comparison is important because $G_0W_0$ band structures are computed on top of the underlying Kohn-Sham DFT results (see, for example, Eq.~\eqref{qpeq}). 
Since discrepancies in the $GW$ band structures computed from different numerical approaches are expected to be larger than those at the DFT level, close agreement among the DFT band structures is a necessary prerequisite for a reliable $GW$ benchmark.
}

\ASH{We show the DFT band structure computed with the PBE exchange-correlation functional~\cite{Perdew1996} and SOC in Fig.~\ref{fig7}, for the aug-SZV-MOLOPT basis set (light settings) and the aug-DZVP-MOLOPT (tight settings), and the band structure computed from  QE (with 8 empty bands). 
We observe excellent agreement between the aug-DZVP-MOLOPT calculation and the QE calculation, with a difference of 18~meV on average between their respective PBE+SOC bandgaps  (see Table \ref{t1}), and also with VASP results with a difference of 10~meV. 
The agreement of the PBE+SOC gap computed with the small aug-SZV-MOLOPT basis is also good, the average deviation is  29~meV to QE and 23~meV to VASP (Table \ref{t1}).
%
%
This shows a good agreement of DFT band structures between each code, which validates their use as a starting point for a $GW$ benchmark. 
Table~\ref{t1} also reports the DFT band gaps without SOC and the SOC splitting at the K-point, demonstrating that our approach yields SOC splittings in good agreement with plane-wave reference calculations.
}
\ASH{For comparison, we also computed the band gaps using the original, non-augmented MOLOPT basis sets~\cite{vande2007} (see Table~\ref{t1}). At the DFT level, the results show good agreement with those from augmented basis sets—except for SZV-MOLOPT—indicating that augmentation is not strictly required for fast convergence of the DFT band gap with respect to basis set size.}
\ASH{To assess the accuracy of the SOC implementation, we report in Table~\ref{t1} the spin-orbit splitting of the valence band maximum at the K-point. \DHP{This splitting is crucial as it is involved in determining the (optical) energy difference between A and B excitons, observed in reflectance and photoluminescence spectra \cite{Heinz2010, Urbaszek2018}. It thus plays a crucial role for valley selective optical excitations.} We  observe good agreement between our perturbative SOC implementation and the fully relativistic implementation in QE, with an average difference of 17~meV between the aug-DZVP-MOLOPT and QE results, as an example. The difference is more significant for the selenium-based TMDs, with 32~meV on average whereas the difference is 2~meV for the sulfur-based TMDs. 
%
The results are similar for the comparison with the VASP SOC-splitting. This finding validates our perturbative SOC treatment from HGH pseudopotentials [Eqs.~\eqref{e31},~\eqref{e58} and  Appendix~\ref{sec:SOC}].}

\RP{We demonstrate the impact of SOC on the band structure by giving in Appendix~\ref{sec:SOC} the comparison between the PBE and PBE+SOC bandstructures for all monolayers (Fig.~\ref{fig7_SOC_compar}).
One can therefore see that the SOC lifts the spin degeneracy and therefore splits the band structure, especially at the K-point as we have already discussed. This leads to a reduction of the band gap with respect to the calculation without SOC, especially in the case of WS$_2$ and WSe$_2$ where the band gap is lowered by 0.3~eV (Table~\ref{t1}).}

\begin{figure*}
    \centering
\includegraphics[scale=0.95]{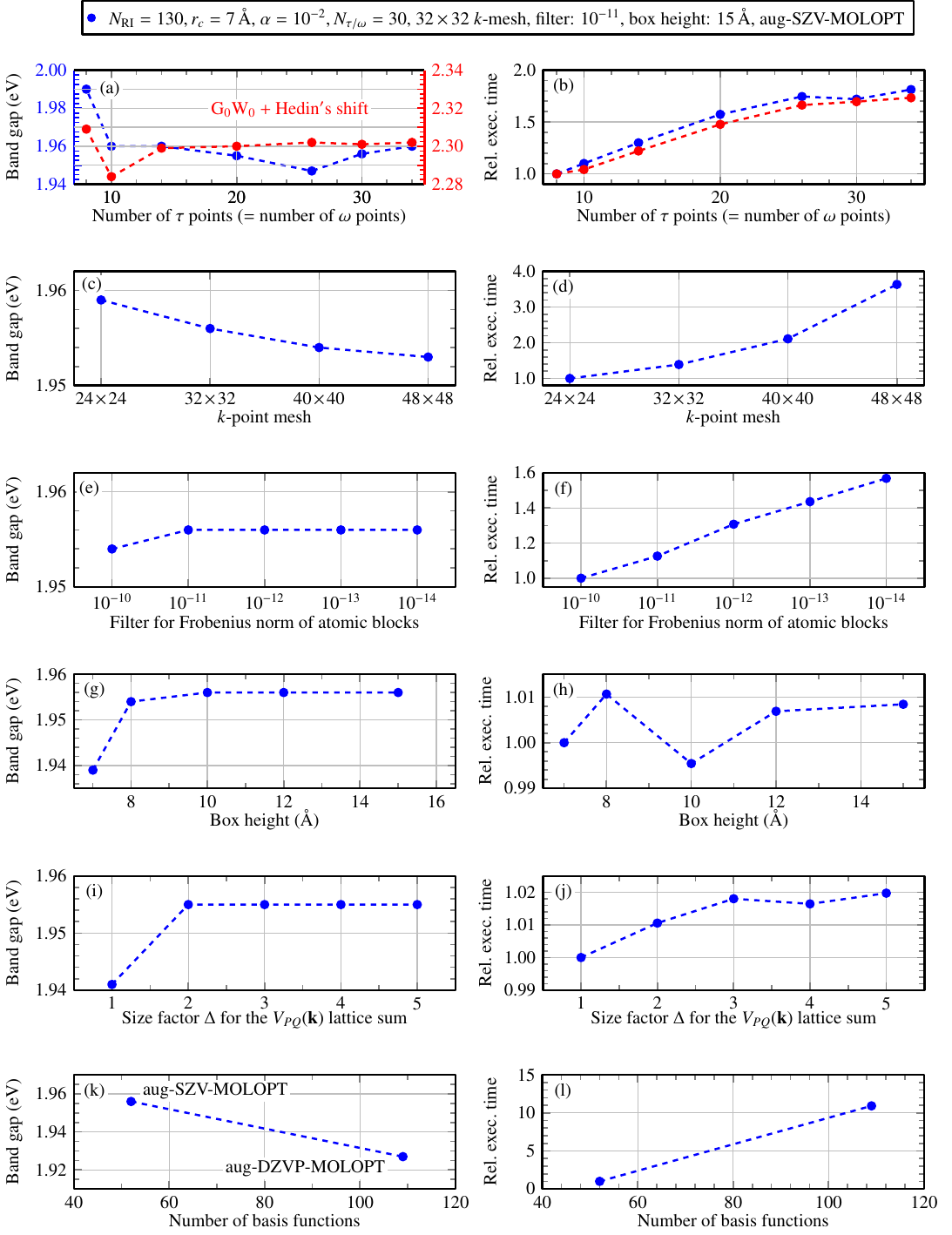}
\vspace{-0.3cm}
    \caption{$G_0W_0$@PBE+SOC band gap of monolayer WSe$_2$ and execution time as a function of the number of time points $\tau$ (Sec.~\ref{sec:IV}), the $k$-mesh  (Eqs.~\eqref{e21a}, \eqref{e23}, \eqref{e29}), the filter threshold (for Eqs.~\eqref{chiT}, \eqref{SigmaR} and~\eqref{SigmaxR}, see Eqs.~\eqref{Frobcut} and \eqref{Frobnorm}), the simulation cell box height (Sec.~\ref{subsec:compparamcp2k}), the size factor $\Delta$ for $V_{PQ}(\mathbf{k})$ (Eq.~\eqref{lattsacle}) and the number of basis functions (Sec.~\ref{subsec:compparamcp2k}). 
    Default parameters are reported on top.
    In (a), we also show $G_0W_0$@PBE+SOC with Hedin's shift~\cite{hedin1965new,Hedin1999,Pollehn1998,Martin_Reining_Ceperley_2016,Golze2019,Golze2022} to avoid poles of the self-energy close to the quasiparticle solution~\cite{Veril2018, Schambeck2024}.
    }
    \label{fig1}
\end{figure*}

\begin{figure*}
\vspace{-0.3cm}
\includegraphics[scale=0.95]{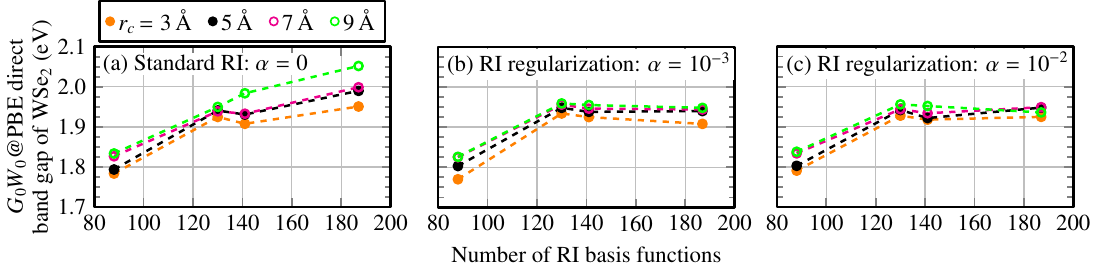}
\vspace{-0.3cm}
    \caption{Direct $G_0W_0$@PBE+SOC band gap of monolayer WSe$_2$ as function of the RI basis set size, the truncation radius~$\rc$ of the truncated Coulomb metric~\eqref{e5c}/\eqref{e11a} and the regularization parameter~$\alpha$, Eq.~\eqref{e12}.
    Tight parameters and the aug-SZV-MOLOPT basis sets are used to expand the KS orbitals. 
    Each point corresponds to an RI basis set with a given relative deviation of the RI-MP2~\cite{Weigend1998} correlation energy compared to MP2: the RI basis set with $N_\text{RI}\eqt 88$ functions gives a relative RI-MP2 error of $10^{-2}$; the other RI basis set sizes and relative RI-MP2 errors are: $N_\text{RI}= 130$ and $10^{-3}$,  $N_\text{RI}= 141$ and $10^{-4}$,   $N_\text{RI}= 187$ and $10^{-5}$.
    }
    \label{fig2}
\end{figure*}

\section{\textit{GW} band gap of WS$\text{e}_\text{2}$: Convergence study}
In this section, we analyze the convergence of the $GW$ bandgap of monolayer WSe$_2$ with the numerical parameters summarized in Sec.~\ref{subsec:compparamcp2k}.
%
%
We focus on the direct band gap of WSe$_2$ at K, calculated using  $G_0W_0$@PBE including SOC. 
\reviewnew{In Appendix~\ref{benchMoS2}, we provide a similar benchmark for MoS$_2$.}

Using our tight convergence settings (Sec.~\ref{subsec:compparamcp2k}), we obtain a $G_0W_0$@PBE+SOC band gap of 1.95~eV with the aug-SZV-MOLOPT basis and 1.93~eV with the aug-DZVP-MOLOPT basis (Table~\ref{t1}). 
A larger augmented basis set is not used due to current memory limitations of our implementation.
Fig.~\ref{fig1}  shows how the band gap changes with various computational parameters.
In Fig.~\ref{fig1}a, we test different sizes of the time-frequency grid. 
We find that the gap changes by up to 12~meV in  $G_0W_0$@PBE (blue trace), when using 14, 20, 26, 30 and 34 points.
These variations can arise by poles in the self-energy~\cite{Schambeck2024}, which is an unphysical feature of  $G_0W_0$@PBE and eigenvalue self-consistency in $G$ will cure this deficiency~\cite{Veril2018,Schambeck2024}. 
We apply Hedin’s shift~\cite{hedin1965new,Hedin1999,Pollehn1998,Martin_Reining_Ceperley_2016,Golze2019,Golze2022} to approximate eigenvalue self-consistency in $G$, which reduces the variation between 14 and 34 time-frequency points to 3~meV (red curve in Fig. 1a).
\RP{The band gap is also converged with respect to the $k$-point mesh from DFT (Eq.~\eqref{kmesh}, Fig.~\ref{fig1}c), changing by less than 10~meV between a $24\timest 24$ and $48\timest 48$ $k$-point meshes.
The filter threshold for three-centre integrals~\eqref{e5c} decides about removing small three-centre integrals from the calculation. 
Thus, decreasing the filter threshold increases the numerical precision and we demonstrate in Fig.~\ref{fig1}e that a filter threshold of $10^{-11}$ is sufficient for sub-meV convergence.
}
\RP{Note that the number of cells included in the lattice sums~\eqref{chiT}, \eqref{SigmaR} and~\eqref{SigmaxR} depend on both the filter threshold of three-centre integrals (Fig.~\ref{fig1}e) and the supercell "SC" defined in Eq.~\eqref{e28}.
The size of the supercell "SC"   is determined by the DFT $k$-mesh [Eq.~\eqref{e28}].
We thus demonstrate convergence w.r.t.~the number of cells in the lattice sums~\eqref{chiT}, \eqref{SigmaR}, and~\eqref{SigmaxR} by demonstrating convergence with the $k$-mesh (Fig.~\ref{fig1}c) and the filter threshold (Fig.~\ref{fig1}e).
 Similarly, we show the fast convergence of the lattice sum~\eqref{Vper} of the Coulomb matrix in Fig.~\ref{fig1}i (the size factor of the lattice summation is defined in Appendix~\ref{app:Vk}).
}
Also, the box height (Fig.~\ref{fig1}g) can be  well-converged.

The computational parameters with the most impact on the computation time are the $k$-mesh (Fig.~\ref{fig1}d) and the basis set (Fig.~\ref{fig1}l).
When applying the present $GW$ algorithm to other materials, we recommend to employ safe numerical parameters for the time-frequency integration (30 points), filter (10$^{-12}$) and box height (15\,\AA), and to focus convergence tests on the $k$-mesh and the basis set.

\begin{figure*}[t!]
    \centering
\includegraphics[width=1.0\textwidth]{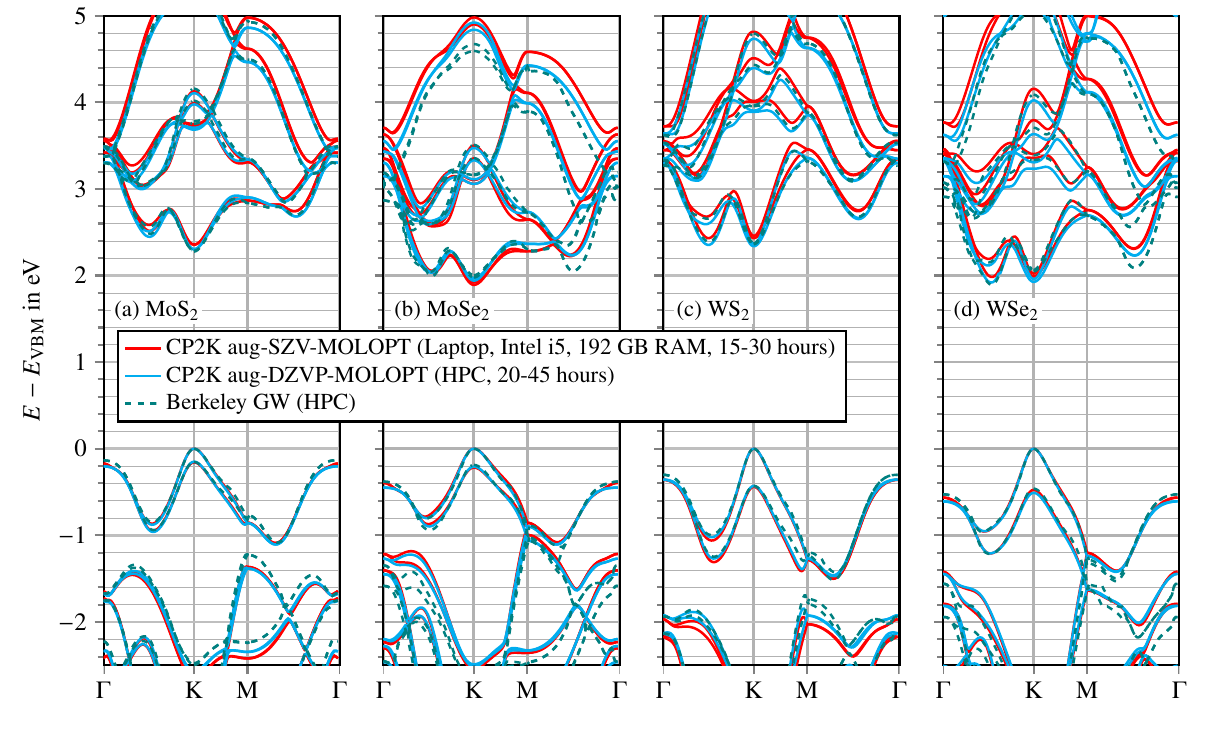}\vspace{-2em}
    \caption{$G_0W_0$@PBE+SOC Bandstructures of monolayer MoS$_2$, MoSe$_2$, WS$_2$ and WSe$_2$, computed from the algorithm presented in this work [CP2K code, Eq.~\eqref{e58}] and computed from BerkeleyGW. 
     The computational details are given in Sec.~\ref{subsec:compparamcp2k} and \ref{subsec:paramQEBGW}.
    }
    \label{fig4}
\end{figure*}

We also study the convergence of the $GW$ gap with respect to the RI basis size in Fig.~\ref{fig2}.
Without regularization $(\alpha\eqt0$), the band gap increases with RI basis size (Fig.~\ref{fig2}a). 
We assign this issue to linear dependencies in the RI basis set, which  lead to numerical instabilities in the RI basis expansion, see Appendix~\ref{app:RI}, Eq.~\eqref{ec2}. 
In a nutshell, numerical instabilities arise when two spatially close diffuse $s$-type RI functions partially compensate each other, leading to large expansion coefficients~$B_{P\bP}^{\mu\bR\nu\bT}$ in Eq.~\eqref{ec2} with alternating sign.
To fix this issue, we use RI regularization~\cite{Graml2024} via $\alpha\eqt10^{-3}$ and $\alpha\eqt10^{-2}$  in Eq.~\eqref{e12}.
RI regularization ensures numerical stability, see Fig.~\ref{fig2}b,c:
For an RI basis set size of a single WSe$_2$ unit between 130 and 187  and a cutoff radius $\rc\intext\{5\,\text{\AA},7\,\text{\AA},9$\,\AA\}, the $GW$ gap is identical within 27~meV.

 \begin{figure}[]
    \centering
\includegraphics[width=0.48\textwidth]{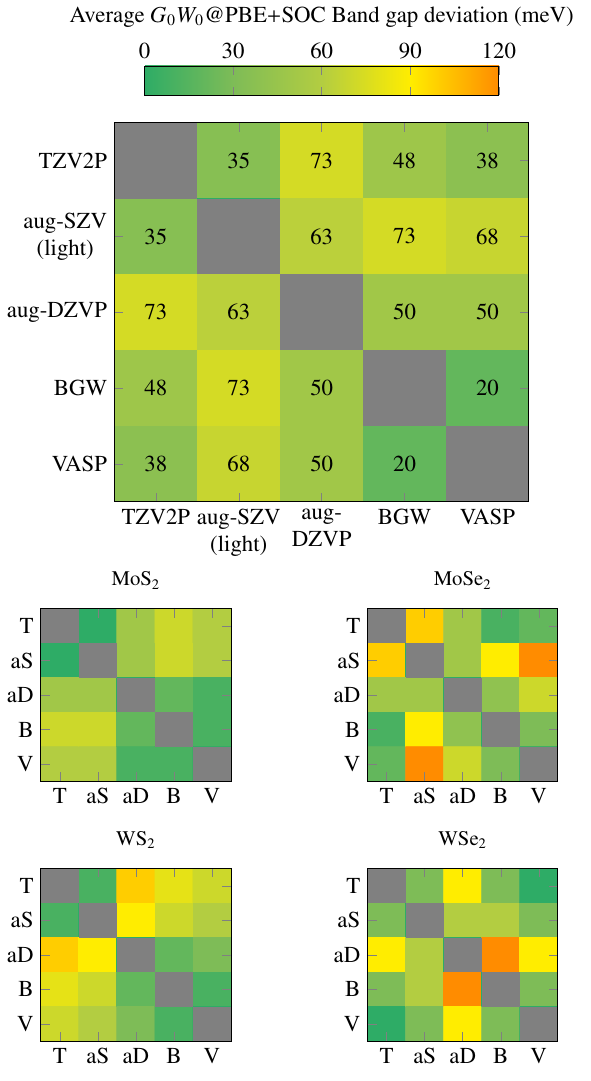}
    \caption{Average absolute deviation in meV across all monolayers TMDs of the $G_0W_0$@PBE+SOC band gap between the TZV2P-MOLOPT (T), aug-SZV-MOLOPT (light, aS), aug-DZVP-MOLOPT (aD), BerkeleyGW (BGW, B) and VASP (V) calculations. The  $G_0W_0$@PBE+SOC band gaps for the individual monolayer TMDs are given \reviewnew{in the four insets below the main figure and }in Table~\ref{t1}.}
    \label{fig_devia}
\end{figure}

\section{\textit{GW} band structure of $\text{MoS}_\text{2}$, 
$\text{MoSe}_\text{2}$, 
WS$_\text{2}$, 
WS$\text{e}_\text{2}$}
\label{sec:GWband}
We employ our $GW$ algorithm to compute the band structure of $\text{MoS}_\text{2}$, 
$\text{MoSe}_\text{2}$, 
WS$_\text{2}$ and
WS$\text{e}_\text{2}$ using two sets of parameters: an aug-SZV-MOLOPT basis with light settings and an aug-DZVP-MOLOPT basis with tight settings, as shown in Fig.~\ref{fig4}.
For comparison, we also include band structures computed using a reference plane-wave-based implementation (BerkeleyGW) with 8 empty bands.
The overall \DHP{qualitative} agreement is \DHP{good}; for quantitative assessment, we focus on the band gap at the K-point (see {Table}~\ref{t1}). 
With the aug-DZVP-MOLOPT basis and tight settings, the $GW$ band gaps agree with the plane-wave results on average within 50~meV. \ASH{The agreement varies across the four materials, and is better for materials containing sulfur (20~meV) than materials containing selenium (75~meV). Note that this discrepancy is also observed between the plane-wave codes (Table~\ref{t1}), as the average deviation is 10~meV for the materials containing sulfur and 30~meV for the materials containing selenium.}
For the aug-SZV-MOLOPT basis and light settings, the average agreement is 70~meV. \ASH{This shows that one can  achieve decent numerical precision of $GW$ band structures using the relatively small aug-SZV-MOLOPT basis set.  
\reviewnew{The observed material dependence of the $GW$ band  gaps can be rationalized by several factors. First, some differences are already present at the DFT level with gaps that differ by 5 meV for MoS\textsubscript{2} and up to 30 meV for WSe\textsubscript{2} (Table~\ref{t1}). The fact that, already at the DFT level with SOC, the differences between the codes are relatively minor in MoS\textsubscript{2} explains, at least in a naïve way, the better agreement among codes after applying the $GW$ correction for this material compared to WSe\textsubscript{2}. Second, the different implementations of SOC in each code (and their corresponding basis sets) can lead to slight differences in the computed dielectric screening, which then propagate to the final $GW$ band gaps. Finally, as noted above, deviations are stronger in selenium-based TMDCs compared to sulfur-based ones. Selenium is more polarizable than sulfur and has more spatially extended valence orbitals, making selenium-based TMDCs more sensitive to how high-energy and local-field effects are represented. }

The average deviations of $GW$ band gaps between the numerical approaches are summarized in Fig.~\ref{fig_devia}.}
\ASH{
The discrepancies may be attributed to several factors, including the use of different pseudopotentials, the limited size of the aug-DZVP-MOLOPT basis compared to high plane-wave cutoffs, and sensitivity to convergence parameters in frequency integration or dielectric matrix evaluation. 
Time-frequency resolution may also contribute to residual differences, which could be mitigated using Hedin’s shift~\cite{Schambeck2024}.
\reviewnew{This is also demonstrated in Appendix~\ref{litbandgap}, where we report some available results in the literature for the band gaps of the TMDCs at the $G_0W_0$@PBE+SOC. As we have stated earlier, one should note however that the comparison with the already existing calculations is fairly difficult as most of these use different lattice constants, which has a huge impact on the band gap~\cite{Zollner2019}.}
}
%

\ASH{
Comparison to experimental measurements is inherently challenging, as the band gap is highly sensitive to external influences that are difficult to control, such as substrate screening effects~\cite{Zhang2016} and strain~\cite{Zollner2019}. Reported experimental band gaps for the four materials range from 1.9~eV to 2.4~eV~\cite{Ryou2016,Zhang2016,Hill2016,Yankowitz2015}, which aligns with our $GW$ results (Table~\ref{t1}). However, achieving a precise one-to-one correspondence for each material remains elusive at this stage, \DHP{especially given that many experimentally reported band gaps are measured in the presence of a substrate, which is not considered in this work, as it lies outside the scope of this manuscript.}
}
%
%

%
{For completeness, we also performed benchmark full-spinor calculations with BerkeleyGW using the generalized plasmon pole model \cite{Hybertsen1986}, see Table \ref{t2} in Appendix \ref{app:GPP}. This approximation yields bandgaps with significant deviations - up to $0.25$~eV -  when compared with the full frequency calculation of the quasi-particle self-energy. These discrepancies sharply contrast with the excellent agreement of below $50$~meV between the different codes, as reported in Fig.~\ref{fig_devia}.}

\begin{figure}
    \centering
\includegraphics[width=0.48\textwidth]{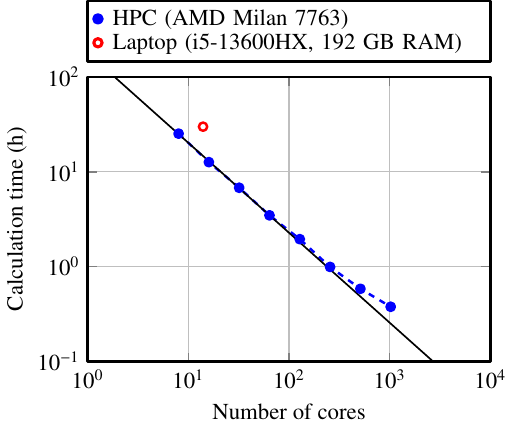}
    \caption{Scaling of the computation time for the $G_0W_0$@PBE+SOC band gap calculation of monolayer MoS$_2$ with respect to the number of cores. 
    We use an aug-SZV-MOLOPT basis and light settings (Sec.~\ref{subsec:compparamcp2k}).
    }
    \label{fig3}
\end{figure}

To demonstrate the computational efficiency of our $GW$ algorithm, we carried out  $GW$ band structure calculations with settings  "aug-SZV-MOLOPT, light" on a laptop with an Intel i5 processor (14 cores) and 192 GB RAM. 
The computation time on this hardware is between 20 and 45 hours, depending on the 2D material. 
\ASH{To study the scalability of our implementation, we ran a computational performance test on an HPC system, using the same parameters as for the laptop calculations in order for the timings to be comparable. The results show that }on this HPC system using 1024 cores, the same calculations completed in approximately 23 minutes (Fig.~\ref{fig3}).
These results demonstrate that our method enables efficient and scalable $GW$ band structure calculations of 2D materials.

\section{Conclusion and outlook} 
We have developed a $GW$ space-time algorithm for periodic systems using Gaussian basis functions with spin-orbit coupling, enabling accurate quasiparticle band structure calculations for atomically thin materials.
Our implementation, available in the open-source code CP2K, achieves high accuracy: for monolayer MoS$_2$, MoSe$_2$, WS$_2$, and WSe$_2$, $GW$ band gaps agree on average within 50~meV to plane-wave $GW$ band gaps from BerkeleyGW and VASP. 
Using lighter computational settings, the average agreement is 70~meV, and a complete $GW$ band structure calculation can be performed on a laptop (Intel i5, 192GB RAM) in approximately 24 hours, or in less than 30 minutes on 1024 HPC cores.
Future extensions of this work will combine the present $GW$ algorithm with real-space grids for $G$ and $W$~\cite{duchemin2019,duchemin2021,Delesma2024}.
We expect this to significantly reduce the number of required lattice summations, further lowering the computational cost.
\ASH{The Gaussian basis sets investigated in this work yield converged band gaps and band structures within $\sim$\,100~meV. Developing Gaussian basis sets that enable convergence of periodic $GW$ calculations ideally within 10~meV is  subject of ongoing work.}

\section*{Data and Code availability}
Inputs and outputs of all calculations reported in this work  are available in a Github repository~\cite{github_repo}.
The algorithm presented in this work is available in the open-source package CP2K~\cite{cp2k,Kuehne2020,Kuehne2025}.

\begin{acknowledgments}
We thank A.~Bussy, M.~A.~García Blázquez, F.~Evers, D.~Golze, J.~Hutter, M.~Krack, M.~Leucke, H.~Pabst, B.~Samal, O.~Schütt for helpful discussions.
We acknowledge  funding by the Free State of Bavaria through the Lighthouse project ”Free-electron states as ultrafast probes for qubit dynamics in solid-state platforms” within the
Munich Quantum Valley initiative,  funding by the Barcelona Supercomputing Center via the 2023 Call for Inno4Scale Innovation Studies, project  Exa4GW, Inno4scale-202301-036, the state of Bavaria  for support via the KONWIHR software initiative and the DFG for funding via the Emmy Noether Programme (project number 503985532), CRC 1277 (project number 314695032, subproject A03) and RTG 2905 (project number 502572516).
{D.H.-P. gratefully acknowledges support from the Diputación Foral de Gipuzkoa through Grants 2023-FELL-000002-01 and 2024-FELL-000009-01, as well as from the Spanish MICIU/AEI/10.13039/501100011033 and FEDER, UE through Project No. PID2023-147324NA-I00 and from IKUR Strategy, Quantum Technologies 2025 project \textit{M-Twist}.} 
{M.C.-G. acknowledges the support from the Diputaci\'on Foral de Gipuzkoa through Grant 2024-FELL-000007-01 and from the Gobierno Vasco-UPV/EHU Project No. IT1569-22.}
The authors gratefully acknowledge the computing time provided to them on the high performance computer Noctua 2 at the NHR Center PC2. These are funded by the Federal Ministry of Education and Research and the state governments participating on the basis of the resolutions of the GWK for the national high-performance computing at universities (www.nhr-verein.de/unsere-partner).
The authors also gratefully acknowledge the Gauss Centre for Supercomputing e.V. (www.gauss-centre.eu) for funding this project by providing computing time on the GCS Supercomputer SuperMUC-NG at Leibniz Supercomputing Centre (www.lrz.de).
D.H.-P. also acknowledges computational resources  provided by the Max Planck Computing and Data Facility (MPCDF). D.H.-P. and M.C.-G. thankfully acknowledge RES resources provided by the Barcelona Supercomputing Center in MareNostrum 5 to FI-2025-1-0014 and the provided technical support.
\end{acknowledgments}

\appendix

\section*{Appendix}

\section{Periodic RI}~\label{app:RI}
For computing~$\chi_{PQ}^\bR(i\tau)$ from Eq.~\eqref{chiT}, the three-centre integrals $(\mu\bR\,\nu\bT |P\bR ) $ 
are appearing. 
Moreover, in Eq.~\eqref{e11a} the metric matrix~$M_{PQ}(\bk)$ is appearing.
$(\mu\bR\,\nu\bT |P\bR ) $ and $M_{PQ}(\bk)$  are stemming from the application of the resolution of the identity (RI)~\cite{Vahtras1993,zhu2021all,Graml2024}. 
In this Appendix, we give details on RI with periodic boundary conditions.
In quantum chemistry, using Gaussian basis functions without periodic boundary conditions, RI is expressed as
\begin{align}
    \phi_\mu(\br)\,\phi_\nu(\br)
    \simeq
    \sum_P B_P^{\mu\nu}\varphi_P(\br)\,,\label{ec1}
\end{align}
where the product of two Gaussian functions~$\phi_\mu(\br)$ and $\phi_\nu(\br)$  is approximated as a linear combination of auxiliary Gaussian functions $\varphi_P(\br)$  with coefficients 
$B_P^{\mu\nu}$. 
$B_P^{\mu\nu}$ are determined by fitting procedures like least-squares minimization~\cite{Vahtras1993} to get a good approximation in Eq.~\eqref{ec1}.
In the periodic case, the basis functions~$\phi_\mu^{\bR}(\br)$ and $\phi^{\bT}_\nu(\br)$ can be located in cell~$\bR$ and $\bT$. 
The RI expansion then uses auxiliary Gaussians~$\varphi_P^\bP(\br)$ in any cell~$\bP$,
\begin{align}
    \phi_\mu^{\bR}(\br)\,\phi_\nu^{\bT}(\br)
    \simeq
    \sum_{P\mathbf{P}} B_{P\bP}^{\mu\bR\nu\bT}\varphi_P^\bP(\br)\,.\label{ec2}
\end{align}
We show in this Appendix that the expansion coefficients~$B_{P\bP}^{\mu\bR\nu\bT}$ are determined by least-squares minimization under a metric~$m(\br)$~\cite{Vahtras1993} as
\begin{align}
B_{P\bP}^{\mu\bR\nu\bT}
=\sum_{Q\bQ} 
(\mu\bR \,\nu\bT\,|\,Q\bQ)_m\,(M^{-1})_{PQ}^{\bP-\bQ}\,,\label{ec3}
\end{align}
where $(\mu\bR \,\nu\bT\,|\,Q\bQ)_m$ is the three-centre integral of the metric~$m(\br)$ ($\bQ$ is a lattice vector),
\begin{align}
    (\mu\bR \,\nu\bT&\,|\,Q\bQ)_m 
=
{\int} d\br\,d\br'\;
\phi_\mu^{\bR}(\mathbf{r})\,
\phi_\nu^{\bT}(\mathbf{r})\,
m(\br-\br') \,
\varphi_Q^{\bQ}(\mathbf{r}') \,,
\end{align}
 and 
\begin{align}
(M^{-1})_{PQ}^{\bR}   = \intbz\; \emikR \,M^{-1}_{PQ}(\mathbf{k})
\label{ec4}
\end{align}
where $M^{-1}_{PQ}(\mathbf{k})$  are matrix elements of the inverse  of the metric matrix
\begin{align}
M_{PQ}(\mathbf{k})  =&
\sum_\mathbf{R}\eikR {\int}  d\mathbf{r}\,d\mathbf{r}'\,
\varphi_P^\bo(\mathbf{r}) \, m(\br-\br') \,\varphi_Q^\mathbf{R}(\mathbf{r}')\,.  
\label{ec5}
\end{align}
Note that we regularize~$\bM^{-1}(\bk)$, Eq.~\eqref{e12}, to prevent for linear dependencies and thus large expansion coefficients~$B_{P\bP}^{\mu\bR\nu\bT}$ with alternating sign in the RI expansion~\eqref{ec2}.

In the following, we outline our proof of Eq.~\eqref{ec3}.
As in non-periodic RI~\cite{Vahtras1993}, we define the residual~$\res$ of Eq.~\eqref{ec2}, 
\begin{align}
\res (\br) = \phi_\mu^{\bR}(\br)\,\phi_\nu^{\bT}(\br)
\mt    \sum_{P\mathbf{P}} B_{P\bP}^{\mu\bR\nu\bT}\varphi_P^\bP(\br)\,.\label{ec7}
\end{align}
Now, we vary the expansion coefficients~$B_{P\bP}^{\mu\bR\nu\bT}$ in Eq.~\eqref{ec7} to minimize the repulsion of $\res$ with itself in the metric~$m$,
\begin{align}
(\res|\res)_m\hspace{-0.1em} = \hspace{-0.2em}
 {\int}  d\mathbf{r}\,d\mathbf{r}'\;
\res(\br)\; m(\br-\br') \;\res(\br')\geq 0\;\rightarrow\;\text{min}\,.
\end{align}
In the ideal case, we have $\res\eqt0$ yielding zero repulsion of $\res$ with itself.
In the general case, we are looking for a minimum of  $(\res|\res)_m$, i.e. we take $\partial (\res|\res)_m /\partial B_{P\bP}^{\mu\bR\nu\bT}\eqt 0$  which gives
\begin{align}
-2 (\mu\bR \,\nu\bT\,|\,P\bP)_m
+2\sum_{Q\bQ} B_{Q\bQ}^{\mu\bR\nu\bT} (Q\bQ|P\bP)_m  
=
0
\end{align}
and thus
\begin{align}
\sum_{Q\bQ} B_{Q\bQ}^{\mu\bR\nu\bT} M_{PQ}^{\bP-\bQ}    
=
(\mu\bR \,\nu\bT\,|\,P\bP)_m\,.\label{ec10}
\end{align}
We insert
\begin{align}
M_{PQ}^{\bP-\bQ}   = \intbz\; e^{-i\bk\cdot(\bP-\bQ)} \,M_{PQ}(\mathbf{k})
\end{align}
into Eq.~\eqref{ec10}, we multiply with $e^{i\bq\cdot \bP}$ and  we sum over all lattice vectors $\bP$:
\begin{align}
\sum_{Q\bQ} B_{Q\bQ}^{\mu\bR\nu\bT}
\intbz\; e^{ i\bk\cdot\bQ}& \,M_{PQ}(\mathbf{k})
\sum_\bP e^{i(\bq-\bk)\cdot\bP}\nonumber
\\ &
= \sum_{\bP} e^{ i\bq\cdot\bP}
(\mu\bR \,\nu\bT\,|\,P\bP)_m\,. 
\end{align}
We use $\sum\limits_\bP e^{i(\bq-\bk)\cdot\bP}\eqt \Omega_\text{BZ}\, \delta(\bq{-}\bk)$, $\int\limits_\text{BZ}d\bk\, f(\bk)\delta(\bq{-}\bk)\eqt f(\bq)$ and we multiply with the matrix $\bM^{-1}(\bq)$ to obtain
\begin{align}
\sum_{\bQ} B_{Q\bQ}^{\mu\bR\nu\bT}  
 e^{i\bq\cdot\bQ}
= \sum_{P\bP} e^{ i\bq\cdot\bP}
(\mu\bR \,\nu\bT\,|\,P\bP)_m\,M^{-1}_{PQ}(\bq)\,. 
\end{align}
We then multiply with $e^{-i\bq\cdot\bQ'}$, we integrate over the BZ ($\bq$) and we use Eq.~\eqref{ec4} as well as
\begin{align}
    \intbzq \,e^{i\bq(\bQ-\bQ')} = \delta_{\bQ\bQ'}
\end{align}
($\bQ$ and $ \bQ'$ are lattice vectors; $\delta_{\bQ\bQ'}$ is the Kronecker-$\delta$) to obtain Eq.~\eqref{ec3}.

\section{Derivation of Eq.~\eqref{chiT} for computing $\chi^\bR_{PQ}(i\tau)$}
\label{sec:derivationchi}
For deriving the irreducible response~$\chi^\bR_{PQ}(i\tau)$ for a periodic system, Eq.~\eqref{chiT}, we start from Eq.~\eqref{e3} and~\eqref{green_spacetime} for computing $\chi$ for a molecule. 
For a molecule, we sum over the molecular quantum numbers $i$ and $a$ for occupied and empty MOs in Eq.~\eqref{green_spacetime}.
For a periodic system, the quantum numbers of the one-electronic states are $i\bk$ and $a\bk$ labeling an occupied and empty band at $k$-point $\bk$ in the BZ. 
We thus replace and abbreviate
\begin{align}
   \sum_i^\text{occ}\hspace{0.7em}\;&\rightarrow\hspace{0.3em}\intbz \;\;\sum_i^\text{occ}\hspace{0.9em} \simeq \hspace{0.7em}\sik 
    \\[0.5em]
\sum_a^\text{empty}\;&\rightarrow\hspace{0.3em}\intbz \;\sum_a^\text{empty} \;\;\simeq \;\sak
\end{align}
so that we get from Eq.~\eqref{e3}/\eqref{green_spacetime} for $\tau\gt0$ omitting the imaginary prefactor:
\begin{align}
 \chi (\br, \br', i \tau) 
 =&\;
-\sik \pik^*(\br)   \pikrp  \expepsik\nonumber
\\
&\times \sak \pakr   \pak^*(\br') \expepsak
\\[0.5em]
\overset{\eqref{kpointsbf},\eqref{e29}}{=}\hspace{-0.5em}
\sum_{\lambda\bR_1\mu\bS_1}&
\left[
\sum_{\bk} e^{i\bk\cdot(\bR_1-\bS_1)} 
G_{\lambda\mu}(-i\tau,\bk) 
\right]
\phi_\mu^{\bS_1}(\br)\,
\phi_\lambda^{\bR_1}(\br')\nonumber
\\
\times 
\sum_{\sigma\bS_2\nu\bR_2}&
\left[
\sum_{\bq} e^{i\bq\cdot(\bR_2-\bS_2)} 
G_{\nu\sigma}(i\tau,\bq) 
\right]
\phi_\nu^{\bR_2}(\br)\,
\phi_\sigma^{\bS_2}(\br')
\\[1em]
\overset{\eqref{e23}}{=}\;\sum_{\lambda\bR_1\mu\bS_1}&
G_{\lambda\mu}^{\bR_1-\bS_1}(-i\tau ) \;
\phi_\mu^{\bS_1}(\br)\,
\phi_\lambda^{\bR_1}(\br')\nonumber
\\
\times 
\sum_{\sigma\bS_2\nu\bR_2}&
G_{\nu\sigma}^{\bR_2-\bS_2}(i\tau ) \;
\phi_\nu^{\bR_2}(\br)\,
\phi_\sigma^{\bS_2}(\br')\,.\label{a5}
\end{align}
We obtain the matrix element~$\chi_{PQ}^\bR(i\tau)$ from the projection of $\chi(\br,\br',i\tau)$ in the RI basis~$\{\varphi_P^\bR(\br)\}$, incorporating the RI metric which is the truncated Coulomb operator~\eqref{e6c}:
\begin{align}
 \chi_{PQ}^\bR(i\tau) =& \braket{\varphi^\bo_P|\chi(i\tau)|\varphi^\bR_Q}
 \\[0.5em]
=& \int d\br_1 \,d\br_2\,d\br_3\,d\br_4\;
 \varphi_P^\bo(\br_1)\;
 \Vtworr(\br_1,\br_2)\;\nonumber
 \\&\hspace{2em}\times 
 \chi(\br_2,\br_3,i\tau)\;
 \Vtworr(\br_3,\br_4)\;
 \varphi_Q^\bR(\br_4)
 \\[1em]
  =& 
 \sum_{\lambda\bR_1}\sum_{\mu\bS_1}
 \sum_{\sigma\bS_2}\sum_{\nu\bR_2}
G_{ \lambda\mu}^{\bR_1-\bS_1}(-i\tau ) \;
G_{\nu\sigma}^{\bR_2-\bS_2}(i\tau )\nonumber
\\[0.5em]
&\hspace{2em}\times 
(\mu\bS_1\,\nu\bR_2\, |\,P\,\bo)\;
(\lambda\bR_1\,\sigma \bS_2 \,|\,Q\bR)\,.
\end{align}
We replace $\bS_1\rightarrowtext\,\bR_1{-}\,\bS_1$ and $\bS_2\rightarrowtext\,\bR_2{-}\,\bS_2$ such that Eq.~\eqref{chiT} follows: 
\begin{align}
 \chi_{PQ}^\bR(i\tau)&
  =
 \sum_{\lambda\bR_1}\sum_{\mu\bS_1}
 \sum_{\sigma\bS_2}\sum_{\nu\bR_2}
G_{ \lambda\mu}^{ \bS_1}(-i\tau ) \;
G_{\nu\sigma}^{ \bS_2}(i\tau )\nonumber
\\[0.5em]
&\times 
(\mu{\bR_1}{-}\bS_1\,\nu\bR_2\, |\,P\,\bo)\;
(\lambda\bR_1\,\sigma \bR_2{-}\bS_2 \,|\,Q\bR)\,.
\end{align}

\begin{figure}[b!]
    \centering
    \includegraphics[width=0.47\textwidth]{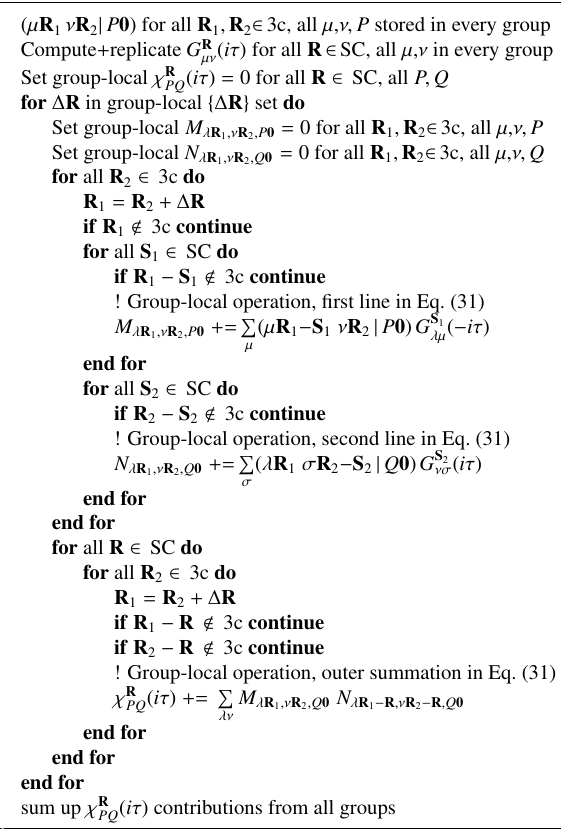}
  \caption{Pseudocode for computing $ \chi^\bR_{PQ}(i\tau)$ from Eq.~\eqref{chiT}.}
  \label{f6}
\end{figure}

\section{Parallel implementation of Eq.~\eqref{chiT} for computing~$ \chi^\bR_{PQ}(i\tau)$}\label{sec:parallelchi}
An efficient, low-memory parallel implementation of Eq.~\eqref{chiT} is a key for routinely executing the present algorithm.
We sketch our parallel implementation in the algorithm from Fig.~\ref{f6}:
The first main idea of the algorithm is to divide the $N$ MPI processes in groups with $n$ MPI processes per group, i.e., the number of groups is $N/n$. 
We store all three-centre matrix elements~$(\mu\bR\,\nu\bS\,|\,P\bo)$ in every subgroup, i.e., the memory available to the subgroup needs to be sufficiently large to fit all $(\mu\bR\,\nu\bS\,|\,P\bo)$ elements. 
This is automatically guaranteed as the program sets the group size $n$ such that 
\begin{align*}
    n \times \text{avail.~mem.~per MPI proc.} > \text{mem.~of }(\mu\bR\,\nu\bS\,|\,P\bo)\,.
\end{align*}
We execute all tensor operations from Eq.~\eqref{chiT} group-locally to avoid communication between all MPI processes.
The second main idea is to distribute the computations of Eq.~\eqref{chiT} on the different groups by distributing $\bR_1{-}\bR_2\,{=}{:}\,\Delta \bR$ to different subgroups. 
It turns out that with this distribution, communication between different groups is avoided; only the group-local result for $\chi^\bR_{PQ}(i\tau)$ needs to be summed up, see last line in the algorithm from Fig.~\ref{f6}.
We employ the set "3c" of lattice vectors $\bR$ which are the lattice vectors~$\bR$ with large three-centre integral~$(\mu\bR\,\nu\bS\,|\,P\bo)$, i.e.,
\begin{align}
\label{Frobcut}
\bR\in\text{3c}
\hspace{0.3em}
\Leftrightarrow
\hspace{0.3em}
\text{there is $\bS$ such that }
\text{F}[(\mu\bR\,\nu\bS\,|\,P\bo)] > \delta\,,
\end{align}
where F denotes the Frobenius norm, 
\begin{align}
\label{Frobnorm}
\text{F}[(\mu\bR\,\nu\bS\,|\,P\bo)]
:=
\sqrt{\sum_{\mu\nu P} |(\mu\bR\,\nu\bS\,|\,P\bo)|^2}\;,
\end{align}
and $\delta$ is the filter threshold. 
All other quantities of the algorithm from Fig.~\ref{f6} are defined in the main text.
The three tensor operations  are executed by the  sparse-tensor library \textit{dbt}~\cite{Seewald2020}.

\section{Lattice sum of the Coulomb matrix element~\eqref{Vper}}
\label{app:Vk}
We compute the lattice sum~\eqref{Vper},
\begin{align}
V_{PQ}(\mathbf{k}) &= 
\sum_\bR e^{i\bk\cdot\bR}\, (\varphi_P^\bo | \varphi_Q^\bR)\;, \label{ed1}
\\
(\varphi_P^\bo | \varphi_Q^\bR) &= 
{\int} d\br\,d\br'\,\varphi_P^\bo(\br)\,\frac{ 1}{|\br- \br'|}\,\varphi_Q^\bR(\br')\,,
\end{align}
by explicit summation of~$\sum_\bR$.
Several schemes for executing this lattice sum have been described, see Ref.~\cite{Piela1982}, Appendix of Ref.~\cite{Grundei2017} and references therein. 
We first note that the lattice summation in Eq.~\eqref{ed1} is divergent if $\varphi_P^\bo $ and $\varphi_Q^\bR$ are $s$-type basis functions in the limit~$\bk\rightarrowtext0$.
This is because  $(\varphi_P^\bo | \varphi_Q^\bR)\simt  |\bR|^{-1}$ and the lattice sum $\sum\limits_\bR (\varphi_P^\bo | \varphi_Q^\bR)\simt \sum\limits_\bR |\bR|^{-1}$ (for $\bk\eqt0$) can be approximated as an integral $\int_{\mathbb{R}^d} d\bR \,|\bR|^{-1}\simt  \int_0^\infty dR \,R^{d-1} \,R^{-1} $ which diverges for dimensionality $d\eqt 1,2,3$ (1d-molecular chains, 2d-materials or surfaces, 3d-bulk solids). 
More precisely, the Coulomb matrix element $V_{PQ}(\mathbf{k})$ diverges at $\bk\rightarrowtext 0$ as $V_{PQ}(\mathbf{k})\simt |\bk|^{-(d-1)}$ for $s$-functions $P,Q$ for  $d\eqt 2,3$ (for $d\eqt1$ the divergence is logarithmic).~\cite{FREYSOLDT20071}
The whole $GW$ algorithm is not divergent because this $1/k^{d-1}$ divergence is integrable in the Brillouin zone, see as an example Eq.~\eqref{trafoWtoR}: $\int_\text{BZ}d\bk \,1/k^{d-1}\simt \int_0^{k_\text{max}} dk\,k^{d-1} / k^{d-1} $, where $k_\text{max}$ is determined by the Brillouin zone size.
The requirements for convergence of the lattice sum~\eqref{ed1} have been extensively discussed in the literature~\cite{Piela1982, Grundei2017} and references therein. 
We reproduce the arguments here to make the discussion self-contained.
We consider the lattice sum 
\begin{align}
    V_{P\rho}(\bk) = \sum_\bT e^{i\bk\cdot\bT}\, (\varphi_P^\bo | \rho^\bT) \label{ed3}
\end{align}
where $\bT$ are lattice vectors and $\rho^\bT(\br)\eqt \rho(\br\mt\bT)$ where $\rho(\br)$ is a function of $\br$ which decays exponentially or faster for large~$|\br|$.
The function $\varphi_P^\bo(\br)$ is assigned to cell $\bo$ and decays exponentially or faster for large $|\br|$.
The lattice sum~\eqref{ed3} is absolutely convergent 
for dimensionality $d\eqt1,2,3$ if
\begin{align}
    k_\text{min} + l_\text{min} \ge d
\end{align}
where $2^{k_\text{min}}$ and $2^{l_\text{min}}$ are the lowest nonvanishing dipole moments of $\varphi^\bo_P$ and~$\rho^\bT$, respectively~\cite{Piela1982}.
In our case, $\varphi^\bo_P$ can be an $s$-type basis function, i.e., $k_\text{min}\eqt 0$, so  
\begin{align}
    l_\text{min} \ge d\label{ed6a}
\end{align}
guarantees absolute convergence of the lattice sum~\eqref{ed3} for all Gaussian basis functions~$\varphi^\bo_P$.
Absolute convergence implies that the order of $\bT$ when executing the lattice sum Eq.~\eqref{ed3} is irrelevant for the result.
Eq.~\eqref{ed6a} means it is required for absolute convergence of  lattice sum~\eqref{ed3} for $d\eqt 1,2,3$ that the function $\rho(\br)$ (and thus $\rho^\bT(\br)$) has zero monopole moment, i.e.,
    \begin{align}
        \int\limits_{\mathbb{R}^3} d\br\;\rho^\bT(\br) =
         \int\limits_{\mathbb{R}^3} d\br\;\rho(\br) =
         0\,.\label{ed6b}
    \end{align}
Additionally, for $d\eqt 2,3$,  all dipole moments of $\rho^\bT$ need to vanish, 
    \begin{align}
        \int\limits_{\mathbb{R}^3} d\br\;r_\alpha\,\rho^\bT(\br) =
         \int\limits_{\mathbb{R}^3} d\br\;r_\alpha\,\rho(\br) =
         0\,.\label{ed6c}
    \end{align}
Additionally, for $d\eqt 3$,  all quadrupole moments of $\rho^\bT$ need to vanish, 
    \begin{align}
        \int\limits_{\mathbb{R}^3} d\br\;r_\alpha\,r_\beta\,\rho^\bT(\br) =
         \int\limits_{\mathbb{R}^3} d\br\;r_\alpha\,r_\beta\,\rho(\br) =
         0\,,\label{ed6d}
    \end{align}
 i.e., Eq.~\eqref{ed6d} needs to hold for all combinations of $\alpha,\beta\eqt1,2,3$.
We apply this theorem to derive an absolutely convergent expression for the lattice sum $V_{PQ}(\bk)$ [Eq.~\eqref{Vper}, reproduced in Eq.~\eqref{ed1}], here illustrated for the case $d\eqt 2$. 
An absolutely convergent lattice summation is essential, as it guarantees that the summation result is independent of the summation order. 
This is particularly important in numerical computations, where the lattice sum must be truncated, thereby imposing a specific summation order.
\begin{figure}
    \centering
    \includegraphics[width=8.6cm]{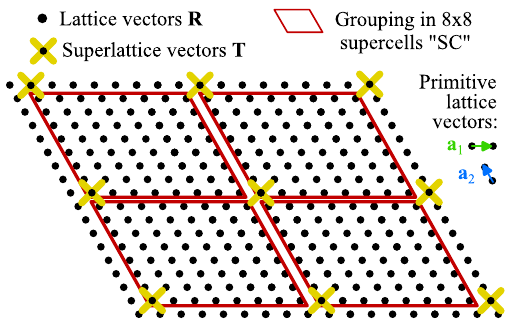}
    \caption{Illustration of lattice vectors~$\bR$ and an examplary 8\,$\times$\,8 supercell SC, defined in Eq.~\eqref{ed5} ($N_1\eqt N_2 \eqt 4$, i.e., the $k$-mesh for BZ integration~\eqref{trafoWtoR} is $4\timest4$).
    }
    \label{fig6}
\end{figure}

We start the derivation of an absolutely convergent lattice summation by defining a $(2N_1)\timest (2N_2)$ supercell "SC" as sketched in Fig.~\ref{fig6}.
Here, $N_1$ and $ N_2$ correspond to the $N_1\timest N_2$ Monkhorst-Pack $k$-point mesh~\cite{Monkhorst1976} used for the BZ integration~\eqref{trafoWtoR} of~$W$.
A lattice vector~$\bR$ belongs to SC if 
 \begin{align}
    \bR\in \text{SC}
    \hspace{0.4em}
    \Leftrightarrow
    \hspace{0.4em}
    \bR = \sum_{j=1}^d n_j\,\ba_j
    \;,\hspace{0.3em}
    n_j\in \{0, 1,\;\ldots\,,\, 2N_j-1\}\,.\label{ed5}
\end{align}
It is then easy to show that the sum of the phase factors~$e^{i\bk\cdot\bR}$ over  the SC lattice vectors   vanishes,
\begin{align}
    \sum_{\bR\in\text{SC}}e^{i\bk\cdot\bR} = 0 \label{ed6}
\end{align}
if $\bk$ is contained in the ${N_1}\timest N_2$ Monkhorst-Pack $k$-point mesh in case $N_1$ and $N_2$ are even integers (see Appendix~\ref{a6a}; in case $N_1,N_2$ are both odd, the $k$-mesh contains the $\Gamma$-point $\bk\eqt 0$ which violates Eq.~\eqref{ed6}).

Eq.~\eqref{ed6} motivates us to reorder the lattice sum~\eqref{ed1} using the superlattice (SL) with lattice  vectors~$\bT$,
\begin{align}
  \bT\in \text{SL}
    \hspace{0.4em}
    \Leftrightarrow
    \hspace{0.4em}
    \bT =  2 N_1\, t_1\,  \ba_1 + 2 N_2\, t_2\,  \ba_2 \;,
    \label{ed7}
\end{align}
where each $t_1, t_2$ is integer.
As illustrated in Fig.~\ref{fig6}, we carry out the infinite lattice sum~\eqref{ed1} by an infinite sum over the superlattice SL:
\begin{align}
V_{PQ}(\mathbf{k}) &=  \sum_{\bT\in\text{SL}} \,
\sum_{\bR\in\text{SC}} e^{i\bk\cdot(\bR+\bT)}\, (\varphi_P^\bo | \varphi_Q^{\bR+\bT}) \label{ed8}
\\[0.5em]
&\equiv \sum_{\bT\in\text{SL}} e^{i\bk\cdot \bT}
\,(\varphi_P^\bo |\rho^\bT) \label{ed9}
\end{align}
where
\begin{align}
   \rho^\bT(\br)= \sum_{\bR\in\text{SC}} e^{i\bk\cdot \bR }\,  \varphi_Q^{\bR+\bT}(\br)\,.
\end{align}
The benefit of lattice sum~\eqref{ed9} is that $\rho^\bT$ has a zero monopole moment, 
    \begin{align}
        \int\limits_{\mathbb{R}^3} d\br\;\rho^\bT(\br) 
     &   =
       \sum_{\bR\in\text{SC}} e^{i\bk\cdot \bR }  \int\limits_{\mathbb{R}^3} d\br\; \varphi_Q^{\bR+\bT}(\br)\nonumber
       \\[0.5em]
       &=
 \sum_{\bR\in\text{SC}} e^{i\bk\cdot \bR }  \int\limits_{\mathbb{R}^3} d\br\; \varphi_Q^{\bo}(\br)
       \overset{\eqref{ed6}}{=}
         0\,, \label{ed11}
    \end{align}
and  zero dipole moments,
   \begin{align}&
        \int\limits_{\mathbb{R}^3} d\br\;r_\alpha\,\rho^\bT(\br) 
        =
       \sum_{\bR\in\text{SC}} e^{i\bk\cdot \bR }  \int\limits_{\mathbb{R}^3} d\br\;r_\alpha\, \varphi_Q^{\bR+\bT}(\br)\nonumber
       \\[0.5em]
       &=
       \sum_{\bR\in\text{SC}} e^{i\bk\cdot \bR }  \int\limits_{\mathbb{R}^3} d\br\;(r_\alpha-R_\alpha-T_\alpha)\, \varphi_Q^{\bo}(\br-\bR-\bT)\nonumber
       \\
       & +
\sum_{\bR\in\text{SC}}  e^{i\bk\cdot \bR }  \,R_\alpha \int\limits_{\mathbb{R}^3} d\br\; \varphi_Q^{\bR+\bT}(\br )
        +
T_\alpha \sum_{\bR\in\text{SC}} e^{i\bk\cdot \bR }  \int\limits_{\mathbb{R}^3} d\br\; \varphi_Q^{\bR+\bT}(\br )
        \nonumber
         \\[0.5em]
       & \overset{\eqref{ed6}}{=}
       0 \,+\,\sum_{\bR\in\text{SC}}  e^{i\bk\cdot \bR }  \,R_\alpha \int\limits_{\mathbb{R}^3} d\br\; \varphi_Q^{\bR+\bT}(\br ) \,+\, 0  = 0\,,
       \label{ed11a}
    \end{align}
where the last equality stems from: 
    \begin{align}
       &\sum_{\bR\in\text{SC}}  e^{i\bk\cdot \bR }  \,R_\alpha=\sum_{(R_\alpha,R_\beta)\in\text{SC}}  e^{i(k_\alpha R_\alpha+k_\beta R_\beta)}\,R_\alpha  \nonumber\\[0.5em]
       &=\sum_{R_\alpha}  e^{ik_\alpha R_\alpha}\,R_\alpha\;\cdot\;\sum_{R_\beta}  e^{ik_\beta R_\beta}  = 0\,,
    \end{align}  
where we used
\begin{align}
    \sum_{R_\beta}  e^{ik_\beta R_\beta} = 0\label{ed17}
\end{align}
as shown in Appendix~\eqref{a6a}, Eq.~\eqref{ae4}. 
 In general, one can easily extend this result and see that for a $d$-dimensional supercell, all moments of order $k\kt d$ vanish if  Eq.~\eqref{ed6} is verified. As such, it is a sufficient condition for the suppression of the quadrupole $(k\eqt2)$ moment for $d\eqt3$, which is condition Eq.~\eqref{ed6d}, and therefore also guarantees the absolute convergence of the three-dimensional lattice sum ~\eqref{ed3}. Comparing Eqs.~\eqref{ed11} and~\eqref{ed11a} to Eqs.~\eqref{ed6c} and~\eqref{ed6d} implies that the lattice sum over~$\bT$ in Eq.~\eqref{ed8} is absolutely convergent, i.e., the result of the lattice sum~\eqref{ed8}/\eqref{ed9} over $\bT$ is independent of the actual order of summation and we thus can truncate  the summation after a finite number of superlattice vectors~$\bT$.
In practice, we check the convergence of the $\bT$ sum by restricting~$\bT$ from Eq.~\eqref{ed7} in lattice sum~\eqref{ed8} to 
\begin{align}
     \bttt =  2 N_1\, t_1\,  \ba_1 + 2 N_2\, t_2\,  \ba_2\;,\hspace{0.5em}t_j\in \{0,\pm1,\ldots,\pm M\}
\end{align}
leading to the lattice sum
\begin{align}
V_{PQ}(\mathbf{k}) &=  \sum_{t_1,\,t_2 = -M}^{M}  \;
\sum_{\bR\in\text{SC}} e^{i\bk\cdot \left(\bR+ \bttt \right)}\; (\varphi_P^\bo | \varphi_Q^{\bR+ \bttt }) \label{ed13}
\end{align}
for  $\bk$ from the even $N_1\timest N_2$ Monkhorst-Pack $k$-point mesh.
Eq.~\eqref{ed13} is implemented in the code and can be converged when increasing $M$. We define the size factor $\Delta$ for Fig.~\ref{fig1} as:
\begin{equation}
    \label{lattsacle}
    M=2^{\Delta-1}
\end{equation}
We  show in Appendix~\ref{appe} that the lattice sum~\eqref{ed13} reproduces the Coulomb integrals used in plane-wave $GW$ algorithms.

\section{Proof of Eq.~\eqref{ed6} and~\eqref{ed17}}
\label{a6a}
 The $N_1\timest N_2$ Monkhorst-Pack $k$-points ($N_1, N_2$ even) are 
\begin{align}
  \bk_\ell   = \sum_{j=1}^{2}  \frac{ \ell_j}{2N_j}\,\bb_j\;,\hspace{1em}
\end{align}
where we define $\ell\eqt(\ell_1,\ell_2)$ and $\ell_j$ takes as value one of the following odd integers
\begin{align}
\ell_j\in \{\pm\,1,\, \pm\,3, \;\ldots\;,\, \pm\, (N_j-1)\}
\,.
\end{align}
$\bb_j$ are  primitive translation vectors of the reciprocal lattice with $\ba_{j_1}\cdott \bb_{j_2}\eqt 2\pi\delta_{j_1j_2}$.  
Then
\begin{align}
  \sum_{\bR\in\text{SC}}\hspace{-0.6em}e^{i\bk\cdot\bR} &\overset{\eqref{ed5}}{=} 
   \sum_{n_1=0}^{2N_1-1} \;\sum_{n_2=0}^{2N_2-1} \hspace{-0.4em} {\exp}\hspace{-0.1em}\left[i
    \left( 
\frac{\ell_1\bb_1}{2N_1} + 
\frac{\ell_2\bb_1 }{2N_2}
     \right)\cdot 
     \left(n_1{\ba_1}{+}\,n_2\ba_2\right)\right] \nonumber
    \\[0.5em]
   &=  
    \sum_{n_1=0}^{2N_1-1} \;\sum_{n_2=0}^{2N_2-1} 
    \exp\left[2\pi i    \left( 
\frac{\ell_1n_1}{2N_1} + 
\frac{\ell_2n_2}{2N_2}
     \right)\right] \nonumber
    \\[0.5em]
   &=  
    \sum_{n_1=0}^{2N_1-1}\left(
    \exp  
\frac{\pi i \ell_1}{N_1}
   \right)^{n_1}
      \;\cdot\;\sum_{n_2=0}^{2N_2-1} 
      \left(
    \exp  
\frac{\pi i \ell_2}{N_2}
    \right)^{n_2}\nonumber
          \\[0.5em]
   &=   
  \frac{1-\left(
    \exp  
\frac{\pi i \ell_1}{N_1}
      \right)^{2N_1}}{1- 
    \exp  
\frac{\pi i \ell_1}{N_1}
      }
      \cdot 
        \frac{1-\left(
    \exp  
\frac{\pi i \ell_2}{N_2}
      \right)^{2N_2}}{1- 
    \exp  
\frac{\pi i \ell_2}{N_2}
      }\nonumber
                \\[0.5em]
   &=   
  \frac{1- 
    \exp   \left(
 2\pi i \ell_1 
      \right)  }{1- 
    \exp  
\frac{\pi i \ell_1}{N_1}
      }
      \cdot 
        \frac{1-
    \exp  
\left(2\pi i \ell_2\right)
    }{1- 
    \exp  
\frac{\pi i \ell_2}{N_2}
     } = 0\,.
\end{align}
In the derivation, we have used $\sum_{n=0}^{N-1}q^n\eqt (1{-}q^N)/(1{-} q)$. In particular, this implies~\eqref{ed17} as for each $\beta=1,2$ direction $k_\beta$ of any of the $\bk_\ell$ with $\ell=(\ell_\alpha,\ell_\beta)$, we have:
\begin{align}
     \sum_{R_\beta}  e^{ik_\beta R_\beta}&= 
   \sum_{n_\beta=0}^{2N_\beta-1} \hspace{-0.4em} {\exp}\hspace{-0.1em}\left[i
    \left( 
\frac{\ell_\beta\bb_\beta}{2N_\beta}
     \right)\cdot 
     \left(n_\beta{\ba_\beta}\right)\right] \nonumber \\
     &=
     \frac{1- 
    \exp   \left(
 2\pi i \ell_\beta 
      \right)  }{1- 
    \exp  
\frac{\pi i \ell_\beta}{N_\beta}
      }=0 \label{ae4}
\end{align}

\section{Proof that absolutely convergent lattice sum~\eqref{ed13} reproduces plane-wave \textit{GW} Coulomb integrals}
\label{appe}
In this Appendix, we prove that the Coulomb matrix computed from the lattice sum from Eq.~\eqref{ed13} gives the same  Coulomb matrix used in plane-wave $GW$. 
We start with the inverse Fourier transform of the Coulomb interaction, 
\begin{align}
   \hspace{-0.15em}{\int\limits_{\mathbb{R}^3}} \frac{d\bG}{(2\pi)^3}&\, \frac{4\pi}{G^2} \,e^{-i\bG(\br-\br')}
    =
    \int\limits_0^\infty \frac{dG}{2\pi^2} \int\limits_0^{2\pi}d\varphi  \int\limits_0^\pi d\theta \sin\theta\,e^{-iG|\br-\br'|\cos\theta}
    \nonumber
    \\[0.1em]
    &=
    \int\limits_0^\infty \frac{dG}{\pi}
    \frac{1}{iG|\br-\br'|}
    \left(e^{iG|\br-\br'| }-e^{-iG|\br-\br'| }        
    \right)\nonumber
    \\[0.1em]
    &=
    \frac{2}{\pi|\br-\br'|}\,\int\limits_0^\infty dG\;
    \frac{\sin (G|\br-\br'|)}{G} = \frac{1}{|\br-\br'|}\,.\label{af1}
\end{align}
For 2D systems, the Coulomb operator is truncated in the non-periodic direction to eliminate artificial image interactions~\cite{Ismail2006} and all following steps are analogous.
We assume an absolutely convergent (a.c.) lattice sum for the Coulomb matrix, like Eq.~\eqref{ed13}, which we abbreviate as
\begin{align}
V_{PQ}(\mathbf{k}) &= 
\sum_\bR^\text{a.c.} e^{i\bk\cdot\bR}\, (\varphi_P^\bo | \varphi_Q^\bR) \,.\label{af2a}
\end{align}
The absolutely convergent lattice summation allows us to interchange the lattice summation and the integration, 
\begin{align}
V_{PQ}(\mathbf{k}) &=  
(\varphi_P^\bo | \sum_\bR^\text{a.c.} e^{i\bk\cdot\bR}\, \varphi_Q^\bR)\,.\label{af2}
\end{align}
Note that if the lattice sum is not absolutely convergent,  we cannot conclude Eq.~\eqref{af2} from Eq.~\eqref{af2a} and the following derivation does not hold.
For connecting the lattice sum~\eqref{af2} to a plane-wave expression, we insert the Fourier transform of the Coulomb interaction~\eqref{af1} into Eq.~\eqref{af2}:
\begin{align}
&V_{PQ}(\mathbf{k}) =  
\int\limits_{\mathbb{R}^3} \int\limits_{\mathbb{R}^3}d\br\,d\br'\,\varphi_P^\bo(\br)\,\frac{ 1}{|\br- \br'|}\,\sum_\bR^\text{a.c.} e^{i\bk\cdot\bR}\,\varphi_Q^\bR(\br')\nonumber
\\[0.2em]
&\overset{\eqref{af1}}{=}
 \int\limits_{\mathbb{R}^3}  \frac{d\bG}{(2\pi)^3}\,\frac{4\pi}{G^2} \int\limits_{\mathbb{R}^3} \int\limits_{\mathbb{R}^3} d\br\,d\br'\,\varphi_P^\bo(\br)\,e^{-i\bG(\br-\br')}\,\sum_\bR^\text{a.c.} e^{i\bk\cdot\bR}\,\varphi_Q^\bR(\br')\nonumber
 \\[0.35em]
&\overset{\eqref{af5}}{=}
\int\limits_{\mathbb{R}^3}  \frac{d\bG}{(2\pi)^3}\,\frac{4\pi}{G^2} \,\varphi_P^*(\bG)\sum_\bR^\text{a.c.} \,e^{i\bk\cdot\bR} {\int\limits_{\mathbb{R}^3} }  \,d\br'\,e^{i\bG\br'}\,\varphi_Q^\bo(\br'-\bR)\nonumber
 \\[0.35em]
&=
\int\limits_{\mathbb{R}^3}  \frac{d\bG}{(2\pi)^3}\,\frac{4\pi}{G^2} \,\varphi_P^*(\bG)\sum_\bR^\text{a.c.} \,e^{i\bk\cdot\bR} {\int\limits_{\mathbb{R}^3} }  \,d\br'\,e^{i\bG(\br'+\bR)}\,\varphi_Q (\br' )\nonumber
 \\[0.35em]
&\overset{\eqref{af5}}{=}
\int\limits_{\mathbb{R}^3}  \frac{d\bG}{(2\pi)^3}\,\frac{4\pi}{G^2} \,\varphi_P^*(\bG)\sum_\bR^\text{a.c.} \,e^{i(\bk+\bG)\cdot\bR}  \varphi_Q (\bG)\nonumber
 \\[0.35em]
&\overset{\eqref{af6}}{=}
\int\limits_{\mathbb{R}^3}  \frac{d\tilde\bG}{(2\pi)^3}\,\frac{4\pi}{\tilde G^2} \,\varphi_P^*(\tilde\bG) \varphi_Q (\tilde \bG)
 \sum_{\bG}^\text{rlv} \Omega_\text{BZ}\,\delta(\bG-(\tilde\bG+\bk))\nonumber
 \\[0.5em]
&\overset{\eqref{af7}}{=}
 \sum_{\bG}^\text{rlv}   \frac{1}{\Omega}\,\frac{4\pi}{ |\bG+\bk|^2} \,\varphi_P^*(\bG+\bk)\, \varphi_Q ( \bG+\bk)\,.\label{af4}
\end{align}
In the course of deriving Eq.~\eqref{af4}, we have used the Fourier transforms of the Gaussian basis functions
\begin{align}
\varphi_P(\bG+\bk)
&= \int\limits_{\mathbb{R}^3} d\br\; e^{i(\bG+\bk)\cdot\br}\,\varphi_P^\bo(\br) 
\,,\label{af5}
\end{align}
the identity
\begin{align}
   \sum_\bR \,e^{i\tilde\bG\cdot\bR}  =
  \sum_{\bG}^\text{rlv} \Omega_\text{BZ}\,\delta(\bG-\tilde\bG)
  \label{af6}
\end{align}
(using the abbreviation rlv for a reciprocal lattice vector~$\bG$; the vector $\tilde\bG$ is arbitrary), and that the Brillouin zone volume~$\Omega_\text{BZ}$ is connected to the unit cell volume~$\Omega$ via
\begin{align}
    \Omega_\text{BZ} = \frac{(2\pi)^3}{\Omega}\label{af7}\,.
\end{align}

Note that Eq.~\eqref{af4} has  been used in Ref.~\cite{Sun2017} to compute the Coulomb matrix element~$V_{PQ}(\bk)$ of Gaussians~$P,Q$.

Eq.~\eqref{af4}  is closely related to the Coulomb matrix in a plane-wave basis, given by
\begin{align}
V_{\bG\bG'}(\bk) = \frac{4\pi}{|\bk+\bG|^2}\,\delta_{\bG\bG'} \,, \label{ed14a}
\end{align}
which arises in plane-wave $GW$ implementations.
Eq.~\eqref{ed14a} and Eq.~\eqref{af4}  are connected via a basis transformation from the plane-wave basis to the Gaussian basis.
This shows that our absolutely convergent lattice summation scheme~\eqref{ed13} yields Coulomb matrix elements fully consistent with~\eqref{ed14a},  ensuring compatibility with plane-wave-based $GW$ implementations.

\section{Derivation of Eq.~\eqref{SigmaR} for computing $\Sigma^{\text{c},\bR}_{\lambda\sigma}(i\tau)$}
\label{sec:derivationsigma}
For the derivation of Eq.~\eqref{SigmaR}, we review
\begin{align*}
\Sigma_{n\bk}(i\tau) &= 
\braket{\psi_{n\bk}|\Sigma(i\tau)|\psi_{n\bk}}
\\[0.5em]&
=
\intcell \int\limits_{\mathbb{R}^3}d\br'\,
  \psi^*_{n\bk} (\br)\,
  \Sigma(\br,\br',i\tau)\,
    \psi_{n\bk} (\br')
\\[0.5em] & =
\sum_{\mu\nu}
C_{\mu n}^*(\bk) \, \Sigma_{\mu\nu}(\bk,i\tau)\,C_{\nu n}(\bk)
\,.
\end{align*}
To arrive at the last line, which is Eq.~\eqref{e17}, we have used the basis expansion~\eqref{kpointsbf} and the $\bk\leftrightarrow\bR$ transformation~\eqref{e48},
\begin{align}
&\Sigma_{\mu\nu}(\bk,i\tau) =
 \sum_\bR^\text{SC} \eikR\;
\Sigma_{\mu\nu}^{\bR}(i\tau) \,.
\end{align}
We further have
\begin{align}
&\Sigma_{\mu\nu}^{\bR}(i\tau)
=
\int d\br \,d\br'\,
\phi_\mu^\bo(\br)\,
G(\br,\br',i\tau) \,
W(\br,\br',i\tau) \,
\phi_\nu^\bR(\br')\nonumber
\\
&\overset{\eqref{a5}}{=} \hspace{-0.8em}
\sum_{\lambda \bS_1,\sigma\bS_2} \hspace{-0.8em}
G_{\lambda\sigma}^{\bS_2-\bS_1}\hspace{-0.3em}
\int d\br \,d\br'\,
\phi_\mu^\bo(\br)
  \phi_\lambda^{\bS_1}(\br)
  W(\br,\br',i\tau)\phi_\sigma^{\bS_2}(\br') 
\phi_\nu^\bR(\br')\,.\label{eqe1}
\end{align}
Inserting periodic RI~\eqref{ec2}/\eqref{ec3} for the products $\phi_\mu^\bo(\br)
  \phi_\lambda^{\bS_1}(\br)$ and $\phi_\sigma^{\bS_2}(\br') 
\phi_\nu^\bR(\br')$ into Eq.~\eqref{eqe1} leads to Eq.~\eqref{SigmaR}.


\section{Spin-orbit coupling from HGH pseudopotentials}\label{sec:SOC}
We employ spin-orbit coupling (SOC) from Hartwigsen-Goedecker-Hutter (HGH) pseudopotentials~\cite{Hartwigsen1998,Vogt2025},
\begin{align}
\hat{V}^{\tmop{SOC}}(\br,\br') &= \sum_l
    \Delta V^\text{SO}_l(\br,\br')\, \frac{\hbar}{2}\,\mathbf{L}\cdot\hat{\boldsymbol{\sigma}}
\\ 
\Delta V^\text{SO}_l(\br,\br')&=\hspace{-0.2em}\sum_{i,j=1}^3\sum_{m=-l}^l \hspace{-0.2em}Y_{lm}(\theta,\varphi)\,p^l_i(r)\,k^l_{ij}\,p^l_j(r')Y_{lm}^*(\theta',\varphi') \label{e2}
\end{align}
where $\hat{V}^{\tmop{SOC}}(\br,\br')$ is the non-local SOC part to the HGH pseudopotential, $\mathbf{L}\eqt{-}i\hbar\,\br\timest\nabla_\br$ the angular momentum, $ \hat{\boldsymbol{\sigma}}$ the Pauli matrices, $Y_{lm}$ the spherical harmonics, $\br\eqt(r,\theta,\varphi)$ are spherical coordinates, $p^l_i(r)$ are tabulated Gaussian functions and $k_{ij}^l$ are tabulated SOC parameters.
We compute SOC matrix elements in the Gaussian basis as
\begin{align}
V_{\mu\nu,\sigma\sigma'}^{\tmop{SOC}}\hspace{-0.2em}(\bk) &=
\sum_\bR \eikR\hspace{-0.2em}{\int} d\br\,d\br'\, \phi_\mu^\bo(\br) \braket{\sigma|\hat{V}^{\tmop{SOC}}(\br,\br')|\sigma'}
\phi_\nu^\bR(\br')
\end{align}
where $\sigma,\sigma'\intext\{\uparrow,\downarrow\}$ is the spin quantum number along the $z$-quantization axis.
We compute the SOC matrix elements in the Bloch basis, 
\begin{align}
    V_{nn',\sigma\sigma'}^{\tmop{SOC}}\hspace{-0.2em}(\bk)
    &=
    \sum_{\mu\nu}
    [C_{\mu n}^\sigma(\bk)]^* \;
    V_{\mu\nu,\sigma\sigma'}^{\tmop{SOC}}\hspace{-0.2em}(\bk)\;
    C_{\nu n'}^{\sigma'}(\bk)\,.
\end{align}

\begin{figure*}[t!]
    \centering
\includegraphics[width=1.0\textwidth]{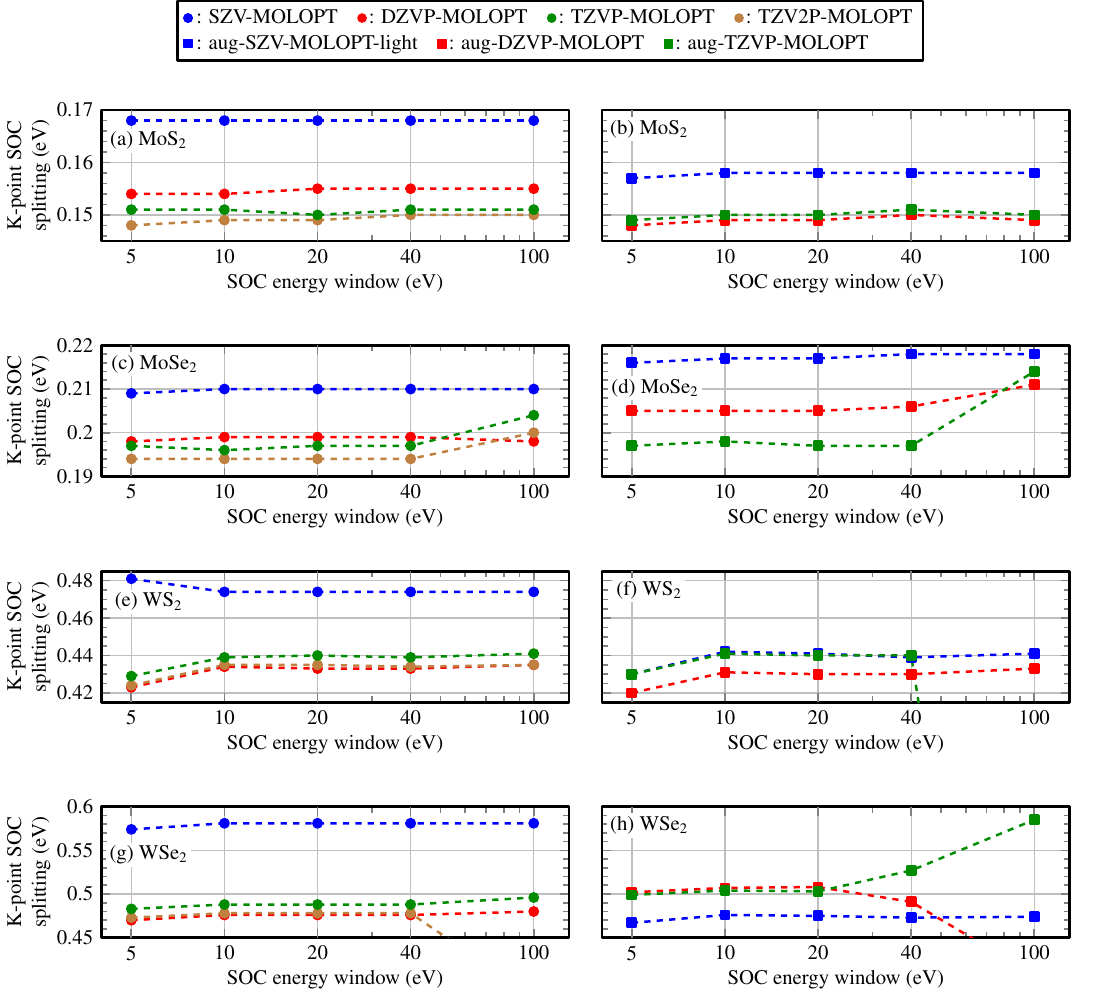}
\vspace{-0.3cm}
    \caption{PBE+SOC K-point spin-orbit splitting as a function of the SOC energy window (Eq.~\eqref{SOC:enerwin}) in monolayer MoS$_2$ for non-augmented MOLOPT (a) and augmented MOLOPT (b) basis sets, in monolayer MoSe$_2$ (respectively (c) and (d)), in monolayer WS$_2$ (respectively (e) and (f)) and in monolayer WSe$_2$ (respectively (g) and (h)).
    The computational details are provided in Sec.~\ref{subsec:compparamcp2k}.
    }
    \label{fig10}
\end{figure*}

\RP{In principle, the SOC matrix~$ V_{nn',\sigma\sigma'}^{\tmop{SOC}}\hspace{-0.2em}(\bk)$ should be computed between all states of the Bloch basis. However, we have observed numerical instabilities originating from nonvanishing pseudopotential overlaps between different atoms. We found that this issue  can be circumvented by restricting the correction to an energy window $E_{\tmop{W}}$ around the valence band maximum~$\varepsilon_{\tmop{VBM}}$ and the conduction band minimum~$\varepsilon_{\tmop{CBM}}$, so that we only include $ V_{nn',\sigma\sigma'}^{\tmop{SOC}}\hspace{-0.2em}(\bk)$ for bands $n,n'$ at $\bk$ that obey
\begin{equation}
\label{SOC:enerwin}
    \varepsilon_{n\mathbf{k}}^{G_0W_0}, \varepsilon_{n'\mathbf{k}}^{G_0W_0}  \in [\varepsilon_{\tmop{VBM}}-E_{\tmop{W}}/2,\varepsilon_{\tmop{CBM}}+E_{\tmop{W}}/2]\,.
\end{equation}
We provide in Fig.~\ref{fig10} the evolution of the K-point splitting in PBE+SOC calculation with respect to this energy window~$E_\text{W}$. We obtain a stable SOC splitting within 1-2~meV for $E_\text{W}\eqt 10$\,eV and $E_\text{W}\eqt 20$\,eV, validating our approach~\eqref{SOC:enerwin}. For WSe$_2$  with aug-DZVP-MOLOPT and aug-TZVP-MOLOPT (Fig.~\ref{fig10}h), we observe deviations of more than 10~meV when increasing the energy window to $E_\text{W}\eqt 40$\,eV and $E_\text{W}\eqt 100$\,eV, so that we chose a window of 20~eV for these calculations, and 40~eV for the other ones for the SOC calculations reported in the main text.}
We build the single-particle Hamiltonian with SOC,
\begin{align}
h^\gwsoc_{n\sigma,\,n'\sigma'}\hspace{-0.1em}(\bk)
=
\delta_{nn'}\,\delta_{\sigma\sigma'}\,\varepsilon_{n\mathbf{k}}^{G_0W_0}
+ 
  V_{nn',\sigma\sigma'}^{\tmop{SOC}}\hspace{-0.2em}(\bk)\,.
\end{align}
We diagonalize $\mathbf{h}^\gwsoc (\bk)$ to obtain  the band structure $\varepsilon^\gwsoc_{j\bk}$ and coefficients~$C_{n \sigma }^{(j)}(\bk)$ in a perturbative manner,
\begin{align}
 \sum_{n'\sigma'} 
 h^\gwsoc_{n\sigma,\,n'\sigma'}\hspace{-0.1em}(\bk)\;
 C_{n'\sigma'}^{(j)}(\bk)
 =
 \varepsilon^\gwsoc_{j\bk} C_{n \sigma }^{(j)}(\bk)\,.
\end{align}

\begin{figure*}
    \centering
\includegraphics[width=1.0\textwidth]{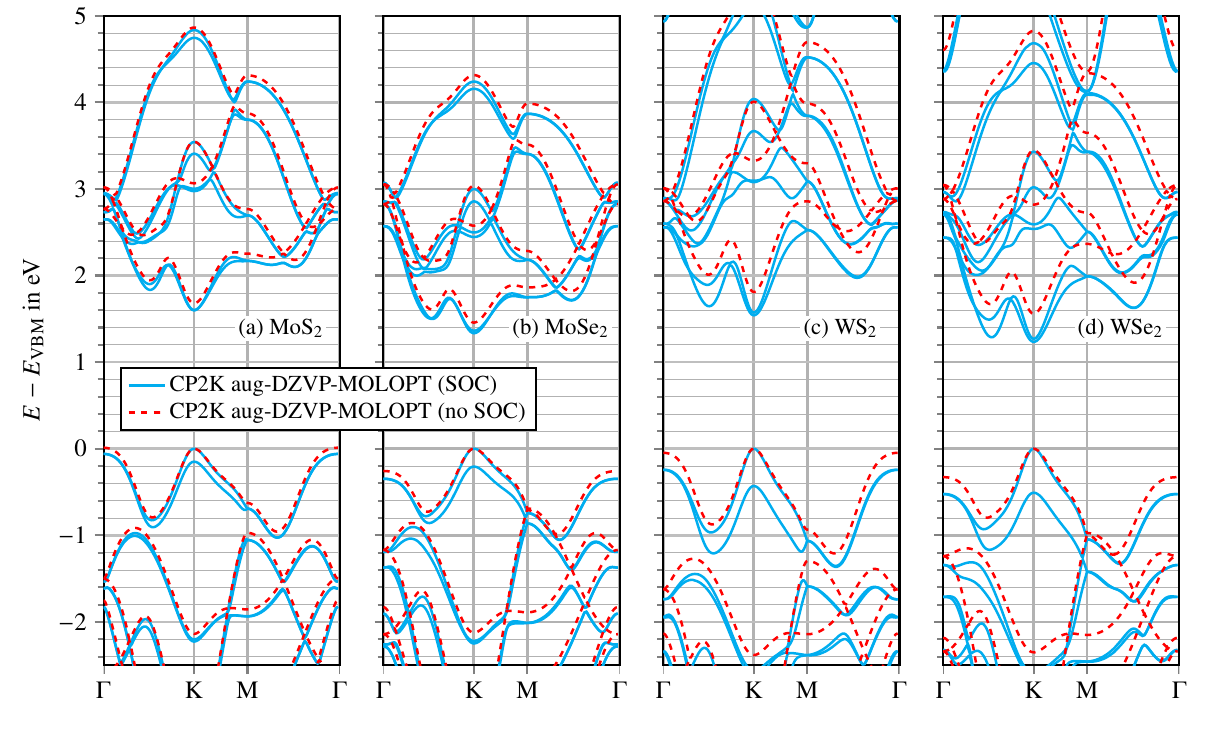}\vspace{-2em}
    \caption{PBE and PBE+SOC Bandstructures of monolayer MoS$_2$, MoSe$_2$, WS$_2$ and WSe$_2$, computed from Eq.~\eqref{e18a} and ~\eqref{e31} using Gaussian basis sets (CP2K).  The computational details are given in Sec.~\ref{subsec:compparamcp2k}.}
    \label{fig7_SOC_compar}
\end{figure*}

\RP{To demonstrate the effects of this correction, we provide in the following the PBE band structures of all monolayers with and without SOC (Fig.~\ref{fig7_SOC_compar}). We observe that SOC splits the top valence bands, predominantly at the K-point, in agreement with the literature~\cite{Gjerding2021}.}

\section{ \textit{GW} band gaps with plasmon-pole model (BerkeleyGW)}\label{app:GPP}
In this Appendix, we present additional data of the band gaps for the TMDCs benchmarked in this work using BerkeleyGW (Table~\ref{t2}). We compare the generalized plasmon pole model (GPP) \cite{Hybertsen1986} with the full frequency implementation (data of the full-frequency calculation is reproduced from Table~\ref{t1}). \reviewnew{Some available results in the literature are given in Table~\ref{t3_GPP}.}

\begin{table}[H]
  \fontsize{10}{12}\selectfont
 \caption{{$G_0W_0$@PBE+SOC bandgap (in eV) using the plane-wave code BerkeleyGW \cite{Hybertsen1986, Deslippe2012, Barker2022} of monolayer $\text{MoS}_\text{2}$, $\text{MoSe}_\text{2}$, WS$_\text{2}$ and WS$\text{e}_\text{2}$. The bandgap is extracted at the K point.}}
{
\renewcommand{\arraystretch}{1.2}
    \begin{tabular}{lcccc}
    \hline
    Calculation & MoS$_2$ & MoSe$_2$ & WS$_2$ & WSe$_2$ 
    \\
    \hline
BerkeleyGW - GPP & 2.53  & 2.12 & 2.53 & 2.13 \\
BerkeleyGW - Full frequency  & 2.28 & 1.98 & 2.36 & 2.05 
      \\
 \hline
    \end{tabular}
}
\label{t2}
 \end{table}

\section{Numerical results on the TMD band gaps in the literature}
\label{litbandgap}
\reviewnew{This section contains a comparison of the results of the paper for the band gaps with the available literature on the topic in Table~\ref{t3_FF} for full frequency calculations and Table~\ref{t3_GPP} for plasmon-pole calculations, staying at the $G_0W_0$@PBE+SOC level. One can easily see a wide range of values, which can be ascribed to the different values for the lattice constants between different calculations, but also to differences in the pseudopotential and SOC calculation method used.} 

\begin{table*}[]
  \fontsize{10}{12}\selectfont
 \caption{Comparison of the $G_0W_0$@PBE+SOC band gaps from this paper and the literature using a full frequency calculation. $L_z$ is the thickness of the vacuum layer (in \AA), $a$ is the lattice constant (in \AA), $N_b$ is the number of bands used in the calculation, $E_{cut}$ is the plane wave cutoff for the DFT calculation and $\varepsilon_{cut}$ is the plane wave cutoff for the calculation of the dielectric tensor.
 }
 \vspace{0.5em}
{\small
\renewcommand{\arraystretch}{1.2}
    \begin{tabular}{lc cccccccc}
    \hline
       Source & Code & Method & $L_z$ (\AA) & a (\AA) & k-grid & N$_b$ & E$_{cut}$ (Ry) & $\epsilon_{cut}$ (Ry) & Direct Gap (eV)
    \\
    \hline
    MoS$_2$&&&&&&&&\\
    \hline
    This work & CP2K & NC (TZV2P) & 15.0 & 3.184 & 32x32x1 & -- & -- & --  & 2.35\\
    This work & CP2K & NC (aug-DZVP) & 15.0 & 3.184 & 32x32x1 & -- & -- & --  & 2.30\\
    This work & Berkeley GW & NC & 15.1 & 3.184 & 12x12x1 & 4000 & 100 & 25  & 2.28\\
    This work & VASP & PAW & 15.1 & 3.184 & 18x18x1 & 384 & 37 & 24.5  & 2.29\\
    C2DB \cite{Haastrup2018,Gjerding2021} & GPAW & PAW & 18.1 & 3.184 & 12x12x1 & N/A & 59 & $\infty$\footnote{\label{note1}extrapolated from 11, 13 and 15 Ry}  & 2.53 \\
      Pela \textit{et al.} \cite{RodriguesPela2024} & exciting & LAPW+lo & 16.0 & 3.190 & 18x18x1 & 400 & N/A & N/A  & 2.45 \\
      Echeverry \textit{et al.} \cite{Echeverry2016} & VASP & PAW & 17.0 & 3.220 & 12x12x1 & 600 & N/A & N/A  & 2.31 \\
      Qiu \textit{et al.} \cite{Qiu2016} & Berkeley GW & NC & 25.0 & 3.150 & 24x24x1 & 6000 & 350 & 35  & 2.45 \\
    \hline
    MoSe$_2$&&&&&&&&\\
    \hline
    This work & CP2K  & NC (TZV2P) & 15.0 & 3.320 & 32x32x1 & -- & -- & --  & 1.99\\
    This work & CP2K  & NC (aug-DZVP) & 15.0 & 3.320 & 32x32x1 & -- & -- & --  & 1.94\\
    This work & Berkeley GW & NC & 15.3 & 3.320 & 12x12x1 & 4000 & 100 & 25  & 1.98\\
    This work & VASP & PAW & 15.3 & 3.320 & 18x18x1 & 384 & 37 & 24.5  & 2.01\\
    C2DB \cite{Haastrup2018,Gjerding2021} & GPAW & PAW & 18.3 & 3.320 & 9x9x1 & N/A & 59 & $\infty$\footref{note1}  & 2.12 \\
      Echeverry \textit{et al.} \cite{Echeverry2016} & VASP & PAW & 17.0 & 3.320 & 12x12x1 & 600 & N/A & N/A  & 2.13 \\
      \hline
    WS$_2$&&&&&&&&\\
    \hline
    This work & CP2K  & NC (TZV2P) & 15.0 & 3.186 & 32x32x1 & -- & -- & --  & 2.44\\
    This work & CP2K  & NC (aug-DZVP) & 15.0 & 3.186 & 32x32x1 & -- & -- & --  & 2.34\\
    This work & Berkeley GW & NC & 18.1 & 3.186 & 12x12x1 & 4000 & 100 & 25  & 2.36\\
    This work & VASP & PAW & 18.1 & 3.186 & 18x18x1 & 384 & 37 & 24.5  & 2.37\\
    C2DB \cite{Haastrup2018,Gjerding2021} & GPAW & PAW & 18.1 & 3.186 & 12x12x1 & N/A & 59 & $\infty$\footref{note1}  & 2.53 \\
    \hline
    WSe$_2$&&&&&&&&\\
    \hline
     This work & CP2K  & NC (TZV2P) & 15.0 & 3.319 & 32x32x1 & -- & -- & --  & 2.02\\
    This work & CP2K  & NC (aug-DZVP) & 15.0 & 3.319 & 32x32x1 & -- & -- & --  & 1.93\\
    This work & Berkeley GW & NC & 15.3 & 3.319 & 12x12x1 & 4000 & 100 & 25  & 2.05\\
    This work & VASP & PAW & 18.3 & 3.319 & 18x18x1 & 384 & 37 & 24.5  & 2.02\\
    C2DB \cite{Haastrup2018,Gjerding2021} & GPAW & PAW & 18.3 & 3.319 & 12x12x1 & N/A & 59 & $\infty$\footref{note1}  & 2.13 \\
    Echeverry \textit{et al.} \cite{Echeverry2016} & VASP & PAW & 17.0 & 3.280 & 12x12x1 & 600 & N/A & N/A  & 2.06 \\
      
 \hline
    \end{tabular}
}
\label{t3_FF}
 \end{table*}

\begin{table*}[]
  \fontsize{10}{12}\selectfont
 \caption{Comparison of the $G_0W_0$@PBE+SOC band gaps from this paper and the literature using a plasmon-pole approximation. $L_z$ is the thickness of the vacuum layer (in \AA), $a$ is the lattice constant (in \AA), $N_b$ is the number of bands used in the calculation, $E_{cut}$ is the plane wave cutoff for the DFT calculation and $\varepsilon_{cut}$ is the plane wave cutoff for the calculation of the dielectric tensor.
 }
 \vspace{0.5em}
{\small
\renewcommand{\arraystretch}{1.2}
    \begin{tabular}{lc cccccccc}
    \hline
       Source & Code & Method & $L_z$ (\AA) & a (\AA) & k-grid & N$_b$ & E$_{cut}$ (Ry) & $\epsilon_{cut}$ (Ry) & Direct Gap (eV)
    \\
    \hline
    MoS$_2$&&&&&&&&\\
    \hline
    This work & Berkeley GW & NC & 15.1 & 3.184 & 12x12x1 & 4000 & 100 & 25  & 2.53\\
      Gillen \textit{et al.} \cite{Roland2017} & Berkeley GW & NC & 25.0 & 3.150 & 24x24x1 & 200 & 88 & 22  & 2.69 \\
      Kim \textit{et al.} \cite{Kim2021} & Berkeley GW & NC & 25.0 & 3.190 & 12x12x1 & 3000 & 125 & 35  & 2.40 \\
      Qiu \textit{et al.} \cite{Qiu2016} & Berkeley GW & NC & 25.0 & 3.150 & 24x24x1 & 6000 & 350 & 35  & 2.59 \\
    \hline
    MoSe$_2$&&&&&&&&\\
    \hline
    This work & Berkeley GW & NC & 15.3 & 3.320 & 12x12x1 & 4000 & 100 & 25  & 2.12\\
      Gillen \textit{et al.} \cite{Roland2017} & Berkeley GW & NC & 25.0 & 3.279 & 24x24x1 & 200 & 88 & 22  & 2.33 \\
      Kim \textit{et al.} \cite{Kim2021} & Berkeley GW & NC & 25.0 & 3.316 & 12x12x1 & 3000 & 125 & 35  & 2.08 \\
      \hline
    WS$_2$&&&&&&&&\\
    \hline
    This work & Berkeley GW & NC & 18.1 & 3.186 & 12x12x1 & 4000 & 100 & 25  & 2.53\\
    Kim \textit{et al.} \cite{Kim2021} & Berkeley GW & NC & 25.0 & 3.189 & 12x12x1 & 3000 & 125 & 35  & 2.46 \\
    \hline
    WSe$_2$&&&&&&&&\\
    \hline
    This work & Berkeley GW & NC & 15.3 & 3.319 & 12x12x1 & 4000 & 100 & 25  & 2.13\\
    Kim \textit{et al.} \cite{Kim2021} & Berkeley GW & NC & 25.0 & 3.334 & 12x12x1 & 3000 & 125 & 35  & 2.01 \\
      
 \hline
    \end{tabular}
}
\label{t3_GPP}
 \end{table*}

 \section{Benchmark study of MoS$_2$}
\label{benchMoS2}
\begin{figure*}
    \centering
\includegraphics[scale=0.95]{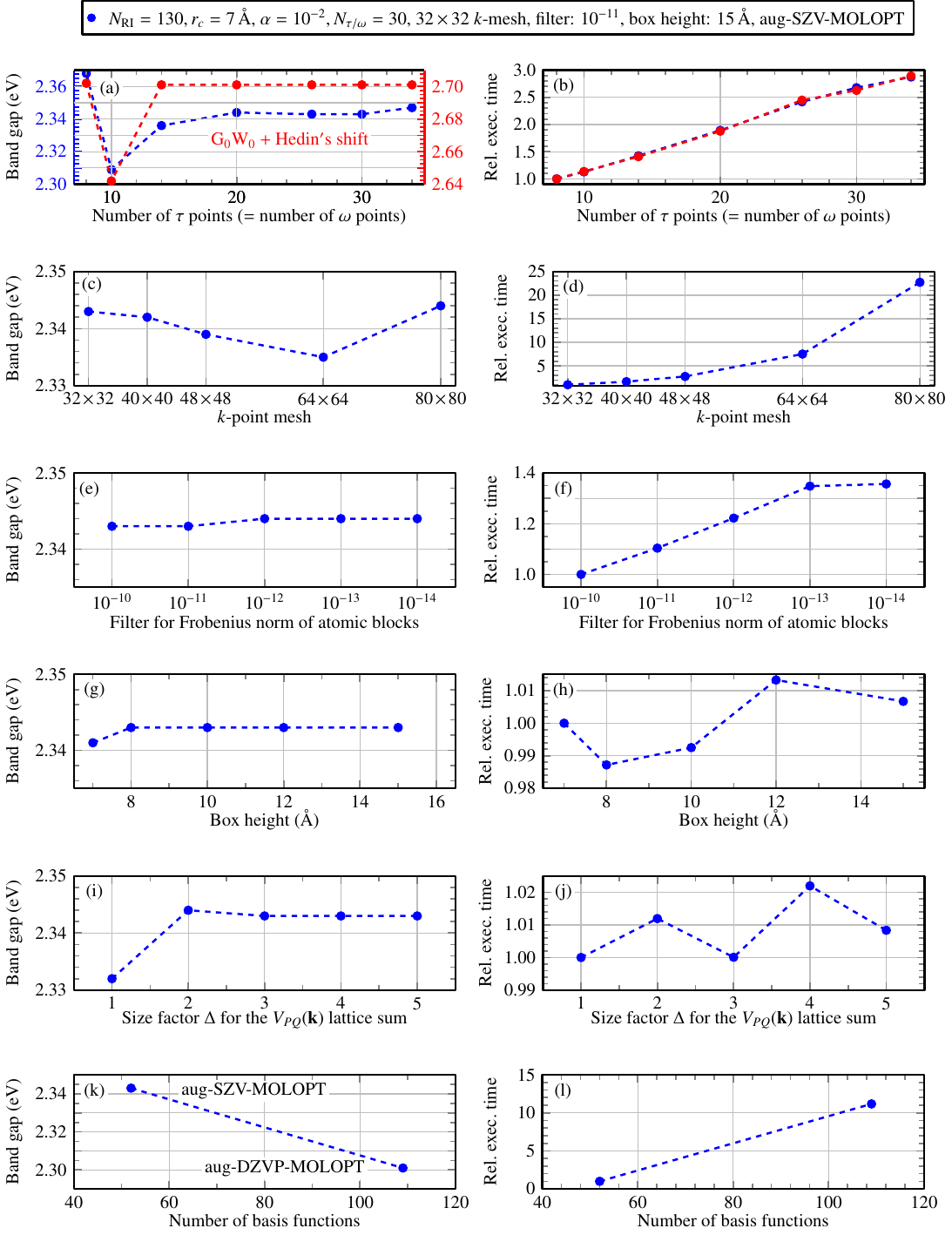}
\vspace{-0.3cm}
    \caption{$G_0W_0$@PBE+SOC band gap of monolayer MoS$_2$ and execution time as a function of the number of time points $\tau$ (Sec.~\ref{sec:IV}), the $k$-mesh  (Eqs.~\eqref{e21a}, \eqref{e23}, \eqref{e29}), the filter threshold (for Eqs.~\eqref{chiT}, \eqref{SigmaR} and~\eqref{SigmaxR}, see Eqs.~\eqref{Frobcut} and \eqref{Frobnorm}), the simulation cell box height (Sec.~\ref{subsec:compparamcp2k}), the size factor $\Delta$ for $V_{PQ}(\mathbf{k})$ (Eq.~\eqref{lattsacle}) and the number of basis functions (Sec.~\ref{subsec:compparamcp2k}). 
    Default parameters are reported on top.
    In (a), we also show $G_0W_0$@PBE+SOC with Hedin's shift~\cite{hedin1965new,Hedin1999,Pollehn1998,Martin_Reining_Ceperley_2016,Golze2019,Golze2022} to avoid poles of the self-energy close to the quasiparticle solution~\cite{Veril2018, Schambeck2024}.
    }
    \label{fig1_MoS2}
\end{figure*}

\reviewnew{In this section, we add a benchmark study of the convergence of the MoS$_2$ band gap, in Fig.~\ref{fig1_MoS2}. As one can see, the convergence behaviour with respect to all parameters is similar to the one observed for the case of WSe$_2$ in Fig.~\ref{fig1}, although the k-point convergence is slightly more unstable (while still being converged within 10~meV)}

\clearpage

\clearpage
\bibliography{main}

\end{document}